\documentclass[iop,apj,12pt,numberedappendix]{emulateapj}
\usepackage{times}
\usepackage[normalem]{ulem}
\usepackage{graphicx}
\usepackage{natbib}
\usepackage{amsfonts}
\usepackage{amsmath}
\usepackage{multirow}
\usepackage{enumerate}
\usepackage{comment}
\usepackage{float}
\usepackage{verbatim}
\usepackage[usenames,dvipsnames]{xcolor}
\usepackage[plainpages=false, colorlinks=true, anchorcolor=blue, linkcolor=blue, citecolor=blue, bookmarks=false]{hyperref}
\newcommand{\be}{\begin{eqnarray}}
\newcommand{\ee}{\end{eqnarray}}

\newcommand{\shortauth}{Morozova et al.}
\newcommand{\slugcom}{Submitted for publication in The Astrophysical Journal}
\slugcomment{\slugcom}

\newcommand{\GG}[1]{}

\lefthead{\sc \footnotesize \slugcom \hfill \shortauth}
\righthead{\sc \footnotesize \slugcom \hfill \shortauth}

\begin{document}

\title{Theoretical X-ray light curves of young SNe~II: the example of SN 2013ej}

\author{Viktoriya Morozova\altaffilmark{1}}
\author{James M. Stone\altaffilmark{1}}
\altaffiltext{1}{Department of Astrophysical Sciences,
  Princeton University, Princeton, NJ 08544, USA, 
  vsg@astro.princeton.edu}

\begin{abstract}

The X-ray signal from hydrogen-rich supernovae (SNe~II) in the first tens to 
hundreds of days after the shock breakout encodes
important information about the circumstellar material (CSM) surrounding their
progenitors before explosion. 
In this study, we describe a way to generate the SN~II X-ray light curves from 
hydrodynamical simulations performed with the code \texttt{Athena++}, 
using the X-ray package \texttt{XSPEC}. In addition, we employ a radiation
diffusion hydrodynamic code \texttt{SNEC} for generating the optical light curves
in different bands. In this numerical setup, we model the X-ray and optical 
emission from a set of progenitor models,
consisting of either two (red supergiant + low density steady wind), or three 
(red supergiant + dense CSM + low density steady wind) components. We vary the
density in the wind and the slope in the CSM to see how these parameters
influence the resulting X-ray and optical light curves. Among our models, 
we identify one that is able to roughly
reproduce both optical and X-ray data of the well observed SN 2013ej. In order to 
achieve this, the slope of the dense CSM in this model should be steeper than
the one of a steady wind ($\rho\propto r^{-2}$), and closer to $\rho\propto r^{-5}$. 
On the other hand, we show that too steep and extended CSM profiles
may produce excessive X-ray emission in the first few tens of days, up to a few orders
of magnitude larger than observed.
We conclude that ability to reproduce the observed X-ray signal from SNe~II together
with their optical light curves is crucial in establishing the validity of different CSM models.

\end{abstract}

	
  
\section{Introduction}

Starting from their birth at zero-age main-sequence (ZAMS), isolated stars
keep loosing mass through their lifetimes, largely in
the form of steady radiation driven winds 
(see reviews of \citealt{kudritzki:00,puls:08,smith:14}). It is 
convenient to estimate these winds in solar masses lost per
year, $\dot{M}$, with the typical values ranging between $10^{-9}$ and 
$10^{-4}\,M_{\odot}\,{\rm yr}^{-1}$, depending on the observational 
diagnostics used to measure them (see \citealt{puls:06,ramirez:17,goldman:17} 
for measurements
in the IR and radio, \citealt{bouret:05,fullerton:06} in the UV, 
\citealt{cohen:10,cohen:11} in the X-rays). 
All modern stellar evolution codes
take these mass losses into account using parameterized prescriptions
for $\dot{M}$ as a function of stellar luminosity, temperature, and metallicity 
(commonly used are the ones of \citealt{jager:88,nieuwenhuijzen:90,vink:01}). 
For example, a $15\,M_{\odot}$ star at ZAMS becomes a red
supergiant (RSG) with the mass slightly less that $13\,M_{\odot}$ by the
end of its evolution \citep{sukhbold:16}. 
Eventually, it undergoes core collapse and explodes,
giving rise to a Type II supernova (SN~II).

In this basic picture, the progenitor of a SN~II 
consists of a RSG surrounded by the wind, the density of which reaches the
values of at most $\sim$$10^{-14}\,{\rm g}\,{\rm cm}^{-3}$ next to the stellar surface and
drops as $\rho\propto r^{-2}$ at larger distances. The influence of such wind
on the optical SN light curve is negligible. For this reason, numerous
hydrodynamical simulations aiming to reproduce the 
observed optical light curves of SNe~II
typically do not include this low density steady wind in the models
\citep[among many others, see the works of][]{bersten:11,
dessart:13,utrobin:17,pumo:17}. At the same time, even a very
low density wind plays a crucial role in the production of X-ray signal from 
SNe~II \citep{chevalier:82,chevalier:16}.

Recently, due to the growing number of early observed, finely sampled SN~II light curves
(\citealt{anderson:14,valenti:16,foerster:16,hicken:17}, see the online database
by \citealt{guillochon:17}), as well as more detailed comparisons to the numerical 
simulations \citep{nagy:16,morozova:17a,paxton:18,moriya:18}, 
we have come to realize that there may be a missing
element in this picture. Agreement between the hydrodynamical models
and observed light curves improves significantly with the addition of a dense compact
circumstellar medium (CSM) in the mass range between $0.1$ and $0.8\,M_{\odot}$,
surrounding the RSG before explosion. It appears to be surprisingly common, seen
in as much as $\sim$$70\%$ of otherwise regular SNe~II 
from the sample studied in \citet{morozova:17a} 
(here, we are not considering SNe~IIn, which are well known for
their strong interaction with CSM, see \citealt{kiewe:12,taddia:13}). 
The formation mechanism of this CSM
is far from being clear. It may be a result of wave-driven enhanced winds or outbursts
in the last stages of stellar evolution \citep{quataert:12,shiode:14,fuller:17}, form during
the common envelope phase with a possible binary companion \citep{chevalier:12}, and
even have a disk-like shape \citep{andrews:17,mcdowell:18}. For many of these scenarios,
it makes little sense to characterize the CSM by the value of $\dot{M}$, 
which implies a steady character of mass losses from the RSG.

Early observations of SNe~II provide numerous hints that could help us
measure the amount and probe the structure of this CSM. For instance, 
many of the early detected SNe~II have narrow features in their spectra,
which disappear within hours of observations (so called `flash spectroscopy'
\citealt{galyam:14,smith:15,khazov:16,yaron:17,hosseinzadeh:18,nakaoka:18,bullivant:18}). 
These observations are
very important in constraining the properties of matter at the site of shock
breakout, and above it. At the same time, numerical models 
of \citet{morozova:17a}, \citet{paxton:18} and \citet{moriya:18}
suggest that the CSM needed to explain
the early (first $\sim$30 days) photometry of SNe~II is so dense, that 
the shock breakout itself happens within this CSM, close to its outer
edge. Therefore, only the outer part of the CSM is flash ionized by the
shock breakout light, while the most part of it cannot be probed by means
of the flash spectroscopy. Another important observational
constraint comes from the rapid rise of SNe~II at the ultraviolet 
wavelengths \citep{gezari:15,tanaka:16,ganot:16}.


In this study, we focus on the relatively early (first $\sim$$150$ days) 
X-ray signal from SNe~II as a potential probe of
the dense CSM surrounding the RSG before explosion. As of now, the
early X-ray data on SNe~II is still scarse, partly because 
of frequent non-detections, which can only place upper limits on the
X-ray luminosity.
Nevertheless, the sample of SNe of all types observed in X-rays has been 
steadily growing and currently consists of over 60 objects 
(\citealt{schlegel:06,immler:07,dwarkadas:12}, see the online database by
\citealt{ross:17}). We have chosen SN 2013ej as an object of our investigation,
because it has 6 X-ray data points in the first 150 days after explosion, as well as
one of the best sampled photometric light curves among SNe~II. We perform
numerical simulations of the SN explosion and generate
X-ray and optical light curves from a set of progenitor models, with and
without dense CSM. Our study shows that
it is possible to construct a model that includes dense CSM and provides a 
good fit for both optical and X-ray data of SN 2013ej. In
previous literature, the influence of the enhanced mass loss in the last years of
stellar evolution on its X-ray light curve was studied, for example, 
by \citet{patnaude:15,patnaude:17}.

The paper is organized in the following way. We start from introducing SN 2013ej 
in Section~\ref{2013ej}. Section~\ref{setup} outlines our numerical setup.
Progenitor models that we use are described in Section~\ref{models}. 
In this study, we work with three numerical codes, \texttt{Athena++}, \texttt{XSPEC}, 
and \texttt{SNEC}, which are described in Sections~\ref{athena}, \ref{xspec}, and \ref{snec},
respectively. Section~\ref{results_low} shows the results obtained from
models with regular low density RSG wind only. Section~\ref{results_CSM} shows the results obtained from
models including dense CSM of a constant slope (one of the examples of which is a
dense steady wind studied in \citealt{morozova:17,morozova:18}). 
Section~\ref{results_Mor} describes the results from models including dense
CSM of a variable slope, similar to the accelerating wind studied in
the works of \citet{moriya:17,moriya:18}. 
Section~\ref{conclusions} is devoted to the conclusions.






\section{Overview of SN 2013ej}
\label{2013ej}

The goal of our study is to present a way of generating SNe~II X-ray light
curves from hydrodynamical simulations, focusing on reproducing the data of SN 2013ej
from a nearby well studied galaxy M74 (NGC 628). 
This SN was discovered very early, less than 1 day after the
last non-detection, by the Lick Observatory Supernova Search 
\citep{dhungana:13,kim:13,shappee:13,valenti:13,lee:13}, and has excellent photometric 
and spectroscopic coverage along the whole duration of its 
light curve \citep{valenti:14,richmond:14,bose:15,huang:15,yuan:16,dhungana:16}. 
Additional important information about SN 2013ej comes from the detection
of its progenitor on the {\it Hubble Space Telescope} (HST) pre-explosion image 
\citep{fraser:14}, search for a possible radio emission \citep{sokolovsky:13}, 
which gave a negative result, and
spectropolarimetric analysis of both the early \citep{leonard:13} 
and late \citep{kumar:16} epochs, 
which showed an unusually strong intrinsic polarization of the SN
\footnote{It should be mentioned that the data of \citet{leonard:13}
were not corrected for the interstellar polarization, but the data of
\citet{kumar:16} were.}.

Semi-analytical studies and numerical simulations of SN 2013ej 
agree on the fact that it had an intermediate mass progenitor.
Among the estimates on the ejected mass of this SN found in the literature 
are $12\,M_{\odot}$ (\citealt{bose:15}, based on the semianalytical 
approach of \citealt{arnett:80,arnett:89,nagy:14}), 
$\sim10.6\,M_{\odot}$ (\citealt{huang:15}, based on the hydrodynamical
simulations), $13.8\pm4.2\,M_{\odot}$ (\citealt{dhungana:16}, based on the
approach of \citealt{litvinova:83}). Modeling the nebular emission lines,
\citet{yuan:16} found the ZAMS mass of the progenitor to be $12-15\,M_{\odot}$.
\citet{fraser:14} gives the range for the progenitor ZAMS mass 
$8-15.5\,M_{\odot}$, based on the analysis of its pre-explosion image. 
On the other hand, hydrodynamical simulations of \citet{utrobin:17}
suggest large ejecta mass of $23-26\,M_{\odot}$.

The classification type of SN 2013ej is somewhat transitional between 
IIP (plateau) and IIL (linear). After the fast initial rise it demonstrated
a fast decline in magnitude (1.74 mag/100 days in $V$-band), and at the
same time a relatively slow decline in the H$\alpha$ and H$\beta$ velocity 
profiles, which is typical for a linear subclass of SNe~II 
\citep[see][]{faran:14a,faran:14b,bose:15}. In \citet{morozova:17}, 
we have shown that in order to reproduce this light curve behavior in 
numerical models, they must include some sort of dense
circumstellar material (CSM) surrounding the RSG before the explosion
\citep[see also][]{das:17}.
The compactness (few thousands of solar radii) and the high
density ($\sim10^{-10}\,{\rm g}\,{\rm cm}^{-3}$) of the 
CSM, which provides the best fit to the data, suggests
that this material could have been ejected from the RSG only a 
few years before the SN explosion. We have reanalyzed SN 2013ej as a part
of a larger SN set in \citet{morozova:17a}, where we found that
the best fitting model for its optical data has ZAMS mass $13\,M_{\odot}$,
final energy $0.68\,{\rm B}$, and $0.49\,M_{\odot}$ of the CSM 
extending up to the radius $1800\,R_{\odot}$. 

It should be mentioned
that SN 2013ej indeed showed some signs of interaction between the
ejecta and CSM, such as an unusually strong absorption feature in the
blue wing of the H$\alpha$ P-Cygni trough\footnote{The nature of this feature
is not entirely clear, and at early times (first $\sim$35 days)
it can be associated with SiII $6355\,${\AA} line \citep{gutierrez:17}.} 
\citep{chugai:07,leonard:13} and presence of
high-velocity components in H$\alpha$ and H$\beta$ profiles \citep{bose:15}.
At the same time, analysis of $\sim$$10$ images of SN 2013ej
host galaxy in $\sim$$5$ years before the explosion suggests 
no significant variability in the progenitor, challenging
the idea of eruptive outbursts in the last few years of stellar evolution,
which could lead to the formation of CSM \citep{johnson:17}.

The X-ray signal from SN 2013ej was first detected with {\it Swift} XRT
\citep{margutti:13} and followed up with {\it Chandra} X-ray Observatory
for five epochs \citep{chakraborti:16}, until day 145 after the explosion. 
 The X-ray data for this and other young
SNe are available online in the Supernova X-ray Database 
(SNaX, \citealt{ross:17}).
The details and analysis of the 
X-ray data are given in \citet{chakraborti:16}, where it was
found that the observations are consistent with the steady
mass loss rate $\dot{M}=2.6\times 10^{-6}\,M_{\odot}\,{\rm yr}^{-1}$
from a $M_{\rm ZAMS}=13.7\,M_{\odot}$ RSG over the last 400 years before
its explosion. In this study we perform the analysis in the inverse direction 
compared to \citet{chakraborti:16}, starting from exploding the 
progenitor model with the wind attached to it, generating the X-ray
signal from the output of the hydrodynamical simulation, and comparing
it to the observational data. Based on the optical modeling suggesting the
presence of high density CSM around the RSG, we attempt to incorporate
the CSM in the X-ray models as well, to obtain a model that would
fit well both optical and X-ray data.


\section{Numerical Setup}
\label{setup}

To model X-ray and optical signal from SN 2013ej we use three
open source numerical codes, \texttt{Athena++} 
\footnote{\url{http://princetonuniversity.github.io/athena/}}, \texttt{XSPEC} 
\footnote{\url{https://heasarc.gsfc.nasa.gov/xanadu/xspec/}}
and \texttt{SNEC} \footnote{\url{https://stellarcollapse.org/SNEC}}, 
applied to the stellar evolution progenitor models
from \texttt{KEPLER} \citep{sukhbold:16}. In this section, we describe the progenitor 
models and numerical setup used in each code. Whenever
possible, we base our choice of numerical parameters either on
the optical light curve fit from \citet{morozova:17a} or on
the choice of parameters used in \citet{chakraborti:16}, in order to facilitate
comparison with their work.

\subsection{Pre-explosion models}
\label{models}

The progenitor models used in this paper are non-rotating solar-metallicity 
red supergiants (RSGs) obtained with the stellar evolution code \texttt{KEPLER} 
\citep{weaver:78,woosley:07,woosley:15,sukhbold:14,sukhbold:16}.
We especially focus on the model with
$M_{\rm ZAMS}=13\,M_{\odot}$, since this model provided the best fit to the
optical data in \citet{morozova:17a}. At the onset of core collapse this 
model has mass $M_{\rm RSG}=11.56\,M_{\odot}$ and radius 
$R_{\rm RSG}=699\,R_{\odot}$. Unless explicitly indicated otherwise,
all X-ray and optical light curves shown in this study are obtained 
from this model.

To generate the X-ray signal from a SN~II it is essential to have a wind 
surrounding the RSG before its explosion. The X-ray emission originates
in a thin shell between the two (forward and reverse) shock waves,
which form after the original, radiation dominated post-explosion shock
wave reaches the surface of the star \citep{chevalier:16,chakraborti:12}.
The X-ray luminosity is expected to depend on the mass loss rate of the
wind $\dot{M}$, its velocity $v_{\rm wind}$ as well as the density slope
in the wind and in the shocked stellar envelope \citep{chevalier:03,dwarkadas:14}.

Observational estimates on the steady mass loss rates from the stars
generally depend on the method used to measure them \citep{smith:14}.
The diagnostics based on measuring free-free emission in radio or IR \citep{wright:75} 
are sensitive to the square of the wind density, $\rho^2$, and predict the
mass loss rates in the range $10^{-7}-10^{-5}\,M_{\odot}\,{\rm yr}^{-1}$. 
Instead, the
diagnostics based on measuring the strength of blueshifted P Cygni 
absorption features \citep{puls:08} 
depend linearly on density and predict the $\dot{M}$ values
in the range $10^{-9}-10^{-6}\,M_{\odot}\,{\rm yr}^{-1}$ 
\citep[see, for example][]{fullerton:06}. It is expected
that the methods measuring $\rho^2$ overestimate the mass loss rates
due to the clumpy structure of the winds 
\citep{owocki:99,bouret:05,dessart:05,sundqvist:13}.

At the same time, growing observational evidence points on the enhanced
mass loss during the last years of stellar evolution, probably in the form of
eruptions \citep[see, for example,][]{mauerhan:13,ofek:14,pastorello:08}. 
In many cases, short signs of interaction with the 
CSM are seen in the early light curves of otherwise regular SNe~II 
\citep{smith:15,khazov:16,yaron:17}.

In the current study, we consider the following three kinds of the pre-explosion 
models:

{\it 1) Models with low density wind only.} We construct these models
by attaching a steady wind with the density profile
\begin{equation}
\label{wind}
	\rho(r) = \frac{\dot{M}_{\rm low}}{4\pi r^2 v_{\rm wind}}
\end{equation}
to the surface of the RSG, and extending it all the way to the outer boundary
of \texttt{Athena++} computational domain ($2.0\times10^{16}\,{\rm cm}$). 
For this kind of models, we take
$v_{\rm wind} = 10\,{\rm km}\,{\rm s}^{-1}$ and use the mass
loss values from the range observed in steady state stellar winds, for example, 
$1-3\times 10^{-6}\,M_{\odot}\,{\rm yr}^{-1}$.

{\it 2) Models including dense CSM with constant slope.} To construct
these models we attach a high density CSM with the density profile
\begin{equation}
\label{CSM}
	\rho(r) = \frac{\mathcal{K}}{r^n}\ ,
\end{equation}
where $n$ and $\mathcal{K}$ are constant, to the surface of the RSG and
extend it up to the radius $R_{\rm CSM}$.
Slope $n=2$ corresponds to the case of a high density steady wind we considered
in \citet{morozova:17,morozova:17a}. In this case, 
$\mathcal{K} = K\equiv \dot{M}_{\rm CSM}/(4\pi v_{\rm CSM})$, where $\dot{M}_{\rm CSM}$
is the effective mass loss and $v_{\rm CSM}$ is the velocity in CSM
(in this work, we do not use $\dot{M}_{\rm CSM}$ anywhere else in the text). 
In \citet{morozova:17a}, we found that the best fitting model for SN 2013ej
had $K=1.0\times 10^{18}\,{\rm g}\,{\rm cm}^{-1}$, 
$R_{\rm CSM} = 1800\,R_{\odot}$, and the total CSM mass 
$M_{\rm CSM} = 0.49\,M_{\odot}$. In this study, we vary the slope
of the CSM, $n$, 
keeping its total mass and radial extent the same as in the best fitting
model from \citet{morozova:17a}. 
For simplicity, we set $v_{\rm CSM} = 10\,{\rm km}\,{\rm s}^{-1}$ everywhere in
this CSM\footnote{Apart from the case $n=2$, assumption of constant velocity
across this CSM does not correspond to any consistent model of steady or 
accelerating wind. However, this has no influence on the light curves obtained from 
these models, which are the main object of our focus.}. 
Above the dense CSM, we attach the regular
low density wind extending all the way to the outer boundary of the domain
($2.0\times10^{16}\,{\rm cm}$).

{\it 3) Models including dense CSM with varying slope.}
Following \citet{moriya:17}, we construct this CSM as an accelerating wind in the form
\begin{equation}
\label{acc}
	\rho(r) = \frac{\dot{M}_{\rm acc}}{4\pi r^2 v_{\rm acc}(r)}\ ,
\end{equation}
with the velocity profile
\begin{equation}
\label{vacc}
	v_{\rm acc}(r) = v_0 + (v_{\infty}-v_0)\left(1-\frac{R_{\rm RSG}}{r}\right)^{\beta}\ .
\end{equation}
Here, $v_0$ is the wind velocity at the stellar surface and $v_{\infty}$ is the 
final wind velocity. We consider the models with $\beta=3.5$, $4.0$, $4.5$, and 
$5.0$, where the last value provided the best fit
to the optical data of SN 2013fs, according to \citet{moriya:17}. Following their work, we 
fix $v_{\infty}$ to be $10\,{\rm km}\,{\rm s}^{-1}$, take 
$\dot{M}_{\rm acc} = 10^{-3}\,M_{\odot}\,{\rm yr}^{-1}$ and place the outer 
boundary of the accelerating wind at $10^{15}\,{\rm cm}$. After that, we adjust
the value of $v_0$ so that the total mass in this CSM is equal to $0.49\,M_{\odot}$,
which we found in \citet{morozova:17a} for SN 2013ej. We justify this step by the fact that
the total mass of the accelerating wind derived in \citet{moriya:17} for SN 2013fs 
is very close to the total mass of the steady wind derived in \citet{morozova:17} for the
same SN. The resulting values of $v_0$ are of the order of $10\,{\rm m}\,{\rm s}^{-1}$.
Above the accelerating wind, we attach the low density wind,
which extends up to the outer boundary of the domain in \texttt{Athena++}
 ($2.0\times10^{16}\,{\rm cm}$).

The model of accelerating wind was proposed in \citet{moriya:17} 
in order to achieve an agreement with the mass loss rate estimate 
of $10^{-3}\,M_{\odot}\,{\rm yr}^{-1}$ made by \citet{yaron:17} for SN 2013fs.
However, the estimate of \citet{yaron:17} itself uses the assumption of a constant velocity
steady wind. Therefore, we do not see a need to try reproducing its value by
a model of accelerating wind. At the same time, a strength of this model
is that the shock breaks out in the outer layers of the dense CSM \citep{moriya:18}, 
flash ionizing a few percent of a solar mass of material in front of it. 
This would be enough to explain the narrow short-living spectral features seen
in SNe~II \citep{dessart:17,boian:17}.

\begin{figure}
  \centering
  \includegraphics[width=0.475\textwidth]{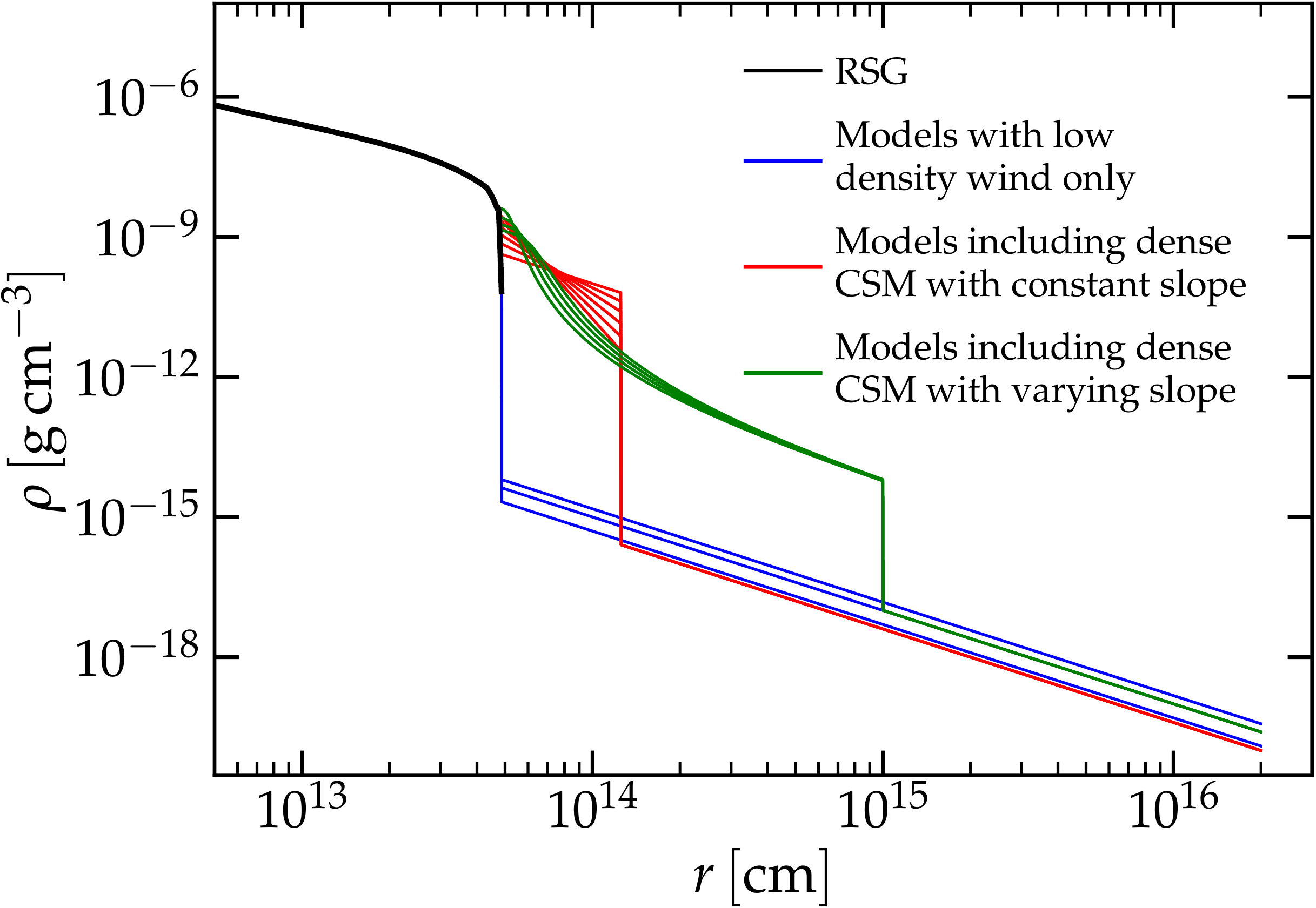}
  \caption{Examples of the density profiles of the three kinds of models
  considered in this study. Black line shows the 
  $M_{\rm ZAMS}=13\,M_{\odot}$ RSG, which is the same
  in all our models.} 
  \label{fig:profiles}
\end{figure}

To summarize, Figure~\ref{fig:profiles} illustrates the density profiles of the
three kinds of pre-explosion models considered in this study.

\subsection{\texttt{Athena++} setup}
\label{athena}

For this study, we use a pure hydro non-relativistic version of the
relativistic magnetohydrodynamic code \texttt{Athena++} 
\citep{gardiner:05,gardiner:08,stone:08}. We assume spherical
symmetry in our models and therefore use 1D version of the code.

\texttt{Athena++} is an Eulerian code and the numerical grid is given
in radial coordinates.
We place our innermost grid zone at the radius, where the density
of the pre-explosion model is equal to $10^{3}\,{\rm g}\,{\rm cm}^{-3}$ 
(radial coordinate $3.6\times10^9\,{\rm cm}$ in our fiducial $13\,M_{\odot}$ model).
This approximately corresponds to the excision of inner $2.4\,M_{\odot}$
of mass. This value is slightly higher than the
remnant masses typically excised in this kind of simulations ($1.4-2\,M_{\odot}$),
but we find that it significantly helps the numerical setup without compromising the
results\footnote{We have performed tests with smaller excision mass and got
very similar results.}. The outermost grid zone is
located at the radial coordinate $2.0\times10^{16}\,{\rm cm}$. The grid consists of
10000 radial points spaced according to a geometric progression with the
ratio between the neighboring cells equal to $1.0011$, so that the inner
region is resolved finer than the outer one. This is necessary in order
to resolve well the underlying RSG model. We note that even with the larger than
usual excision mass the difference in density between the inner and the outer 
grid boundaries in our initial models is larger than 20 orders of magnitude. For the
inner boundary we use a reflective boundary condition, and for the
outer boundary we use an outflowing boundary condition.

We use adiabatic equation of state with the heat capacity ratio $\gamma=1.67$.
In \texttt{Athena++}, as in \texttt{SNEC} (described below), we use a thermal bomb
mechanism to explode the model. The explosion is initiated by injecting the 
internal energy into the innermost $20$ grid points over the time duration of 
$1\,{\rm s}$. Since \texttt{Athena++} simulations do not account for self gravity,
we omit the gravitational energy of the pre-explosion models
when calculating the thermal bomb energy. Initial kinetic and internal energy
of the models after the remnant excision is of the order of $0.03\,{\rm B}$,
which is
small compared to the typical final energies of SNe~II ($0.5-1\,{\rm B}$).
Therefore, in these simulations we take the thermal bomb energy equal to
the final energy we want to reach, $E_{\rm fin}$.
This allows us to put \texttt{Athena++} 
simulations in correspondence with the \texttt{SNEC} simulations of the
same final energy. The simulations are followed for 150 days after the explosion.


By default, \texttt{Athena++} simulations presented here do not include radiation.
However, in order to probe the influence of the radiative cooling on our results,
we perform tests including a calibrated cooling function, as described below in 
Sections~\ref{sec_cooling} and~\ref{results_Mor}.

\subsection{\texttt{XSPEC} setup}
\label{xspec}

In order to generate the X-ray light curve, we analyze the output 
of \texttt{Athena++} simulations using the X-ray package 
\texttt{XSPEC} \citep{arnaud:96}. 

We use the metallicity value $Z=0.295\,Z_{\odot}$, which was 
obtained in \citet{chakraborti:16} based
on the analysis of two-dimensional metallicity distribution
of NGC 628 from \citet{cedres:12}\footnote{The
KEPLER progenitor models that we use were obtained for the
solar metallicity $Z_{\odot}$. However, the difference in the pre-explosion
stellar structure between the metallicities $Z_{\odot}$ and $0.295\,Z_{\odot}$ is 
expected to be small, unlike the difference in the X-ray light
curves.}. The relative metal abundances
are set according to \citet{asplund:09}. The resulting composition
has mean atomic weight $\mu_A=1.24\,{\rm amu}$. 
Assuming complete ionization, this corresponds 
to the value of mean mass per particle $\mu = 0.60\,{\rm amu}$ and
the mean charge state averaged over all ions $q_A=\mu_A/\mu-1=1.07$.
Using this value for $\mu$, we compute plasma temperature
in each grid cell as $T = \mu P/\rho k_{\rm B}$, where $P$ is the
pressure, $\rho$ is the density and $k_{\rm B}$ is the Boltzmann 
constant. The electron and ion number densities are computed
as $n_e=q_A \rho/\mu_A$ and $n_i=\rho/\mu_A$, respectively.

In \texttt{XSPEC}, we use \texttt{tbabs(apec)} model in all grid points
where it is applicable ($0.008<k_{B}T<64\,{\rm keV}$). The APEC
model includes both line and continuum emission from a hot,
optically thin plasma (\citealt{smith:01}, see also 
\url{http://atomdb.org}). In the regions with higher temperatures
($64\le k_{B}T<200\,{\rm keV}$) we use \texttt{tbabs(bremss)}
model. The regions with temperatures higher than 
$200\,{\rm keV}$ are not expected to influence the 
observed part of the spectrum. Following \citet{chakraborti:16}, 
we adopt the value 
$d=9.57\pm0.7\,{\rm Mpc}$ for the distance to the host galaxy
of SN 2013ej, and the reshift value of $z=2.192\times 10^{-3}$ from
NED.


The assumption of complete ionization for computing the ion and
electron number densities in the shocked material
is expected to work well in young supernova remnants, like the one
considered in this study. Since the shock velocities at this stage
are very high ($>500\,{\rm km}\,{\rm s}^{-1}$), the pre-shock matter
is already largely ionized by the photons coming from the shock 
\citep{nymark:06,fransson:84,chevalier:94}.
The recombination time in the shocked ejecta is generally longer than
the cooling time \citep{nymark:06}, leading to some highly ionized
species being present at lower temperatures than they would be
in the steady state case. Furthermore, as will be shown below, 
the relevant temperatures at these times stay very close to 
$\sim10^7\,{\rm K}$, which roughly corresponds to the minimum of the
radiative cooling function for the solar composition (\citealt{nymark:06},
see also Eq.~\ref{cooling} below).
Finally, since the value of metallicity we use for SN 2013ej is smaller
than solar, even incomplete ionization of heavy metals would not
change the electron and ion number densities with respect to the
fully ionized ones in a significant way. However, we emphasize that
this reasoning applies only in the early phase of a supernova remnant
(first few hundreds of days), and in the case of older shocks it is important
to solve for the non-equilibrium ionization state of the ejecta
\citep[see, for example,][]{borkowski:94,borkowski:01,dwarkadas:10}.

Another important assumption that we use in our study is the
equality between the ion and electron temperatures in the shocked ejecta. 
It is known that this assumption is not satisfied in the region
behind the forward circumstellar shock, because the timescale for 
equipartition between the electrons and ions 
\begin{equation}
\label{equipartition}
t_{\rm eq}\approx 2.5\times 10^7 \left(\frac{T_e}{10^9\,{\rm K}}\right)^{1.5}
\left(\frac{n_e}{10^7\,{\rm cm}^{-3}}\right)^{-1}\,{\rm s}\ ,
\end{equation}
where $T_e$ is the electron temperature, becomes too long due to
the high temperature and
low density of the plasma \citep{borkowski:01,chevalier:03,chevalier:16}.
The reverse shock is expected to be in marginal equipartition.
However, as will be seen from the results of the next section,
the main part of the X-ray signal in our models comes from the dense and 
relatively cool shell between the
forward and the reverse shock waves, where the temperatures
of ions and electrons are expected to be equal.

To model the absorption of the X-ray signal with the multiplicative
\texttt{tbabs} model, we compute the hydrogen column at each 
grid cell as $\int_r^{R_{\rm ext}}n_H dr$, where $r$ is the radial coordinate
of the cell
and $R_{\rm ext}$ is the outer boundary of the domain. 
When computing the hydrogen number density $n_H$, we use the
hydrogen mass fraction of $X=0.748$\footnote{In the photospheric 
solar composition
of \citet{asplund:09} ($Z_{\odot}=0.0134$) $X_{\odot}=0.738$, and we
add the value $(1-0.295)Z_{\odot}$ to the mass fraction of hydrogen.}.
To the obtained value of hydrogen column 
we add the constant value for the external absorption column
$n_{\rm ext} = 4.8\times 10^{20}\,{\rm atoms}\,{\rm cm}^{-2}$. This value was 
used in \citet{chakraborti:16}, based on the Leiden Argentine
Bonn (LAB) Survey of Galactic $H_{\rm I}$ \citep{kalberla:05}.

Further in the text, by total X-ray flux we mean the flux 
calculated between $0.5$ and $8.0\,{\rm keV}$, while the soft and
the hard fluxes correspond to the energy ranges $0.5-2.0\,{\rm keV}$
and $2.0-8.0\,{\rm keV}$, respectively. 


\subsection{SNEC setup}
\label{snec}

The optical ($V$-band) synthetic light curves shown in this study were
obtained with \texttt{SNEC} \citep{morozova:15}, a Lagrangian code that 
solves for the hydrodynamics and equilibrium-diffusion radiation transport 
in the expanding envelopes of core-collapse SNe, taking into account
recombination effects and the presence of radioactive nickel.

As in \citet{morozova:17a}, we excise the inner part of the RSG models 
at the mass coordinate of silicon/oxygen interface in their
composition profiles, because it approximately corresponds to the
point of shock revival in the core-collapse explosion mechanism
simulations \citep[e.g.,][]{mueller:12,summa:16,suwa:16,burrows:16,radice:17}. 
For a $13\,M_{\odot}$ model this interface is located at the mass coordinate
$\approx 1.6\,M_{\odot}$.
We assume that all matter deeper than that collapses to form a neutron star.
After the remnant is excised, we initiate the explosion
by injecting the internal energy $E_{\rm bomb} = E_{\rm fin} - E_{\rm init}$,
where $E_{\rm fin}$ is the desired final energy and $E_{\rm init}$ is
the total initial (negative, mostly gravitational) energy,
into the inner $0.02\,M_{\odot}$ of the model for a duration of $1\,{\rm s}$.

\texttt{SNEC} uses the equation of state by \citet{paczynski:83}. The ionization
fractions of hydrogen and helium needed for the equation of state are
found following the approach of \citet{zaghloul:00}.
The numerical grid consists of 1000 mass cells, with the finer resolution
in the interior, where the explosion is initiated, and especially fine resolution
towards the outer boundary, in order to resolve the photosphere at
early times \citep{morozova:15,morozova:16,morozova:17}. 
In \texttt{SNEC} simulations we omit the low density wind, 
because it introduces unnecessary numerical challenge in the form of a
very small timestep. From the previous works 
\citep[see, for example,][]{moriya:11} we know that the wind with 
mass loss rates $<10^{-5}\,M_{\odot}\,{\rm yr}^{-1}$ has a negligible 
effect on the optical light curve. Therefore, the outer boundary of \texttt{SNEC}
grid coincides either with the RSG surface, or with the external radius of the high
density wind, if it is included in the model.

The Rosseland mean opacity is computed from OPAL Type II opacity tables
\citep{iglesias:96}
at high temperatures ($T>10^{4.5}\,{\rm K}$) and from the tables of 
\citet{ferguson:05} at low temperatures ($10^{2.7}<T<10^{4.5}$).
In \texttt{SNEC} we use the opacity floor, which is meant to account for the
line broadening in the rapidly expanding stellar envelope as well as 
possible non-thermal ionization and excitation by gamma-rays, the effects that are not
included in the opacity tables \citep[see, for example,][]{blinnikov:96,bersten:11}.
The value of the opacity floor in our simulations is computed from the
compositional profile and equal to 
$0.01\,{\rm cm}^2\,{\rm g}^{-1}$ for the regions with solar metallicity
$Z=0.02$, $0.24\,{\rm cm}^2\,{\rm g}^{-1}$ for $Z=1$, with the linear 
dependence in between. Before the explosion we smoothen the
composition profiles by passing a ``boxcar" with a width of 
$0.4\,M_{\odot}$ through the models four times. The amount of
radioactive $^{56}$Ni in SN 2013ej is taken to be $0.0207\,M_{\odot}$
\citep{valenti:16}.

Photometric light curves
are computed assuming black body emission and using the \texttt{MATLAB}
package for astronomy and astrophysics for
calculating specific wave bands \citep{ofek:14b}. In \texttt{SNEC}, the radiation 
temperature is assumed to be equal to the effective temperature 
$T_{\rm eff}=\left(L/4\pi\sigma_{\rm SB}R_{\rm ph}^2\right)^{1/4}$, where
$L$ is the bolometric luminosity, $\sigma_{\rm SB}$ is the Stefan-Boltzmann
constant and $R_{\rm ph}$ is the photospheric radius corresponding 
to the optical depth $\tau=2/3$. One of the caveats of
this approach is that it assumes the equality between the temperatures of matter and 
radiation at the photosphere location, which may not be true for
some models, including explosions of RSGs. 
In reality, the opacity in the regions below the photosphere is
dominated by scattering, while the radiation color is determined by deeper layers,
where the absorption opacity is still large enough 
(so called, `color shell', \citealt{nakar:10}). For this reason, the photometric light
curves returned by \texttt{SNEC} generally rise faster than the light curves 
obtained from more sophisticated multi-group radiation-hydrodynamics codes 
(for example, \texttt{STELLA}, \citealt{paxton:18}). If anything, this means
that it is even more challenging for those codes to reproduce the fast rising 
light curves of SNe~II with bare RSG models without CSM. For this reason,
the amount of CSM needed to explain the observed light curves with those 
codes is comparable, and in some cases larger than the one we need in
\texttt{SNEC} \citep[see, for example,][]{moriya:17,paxton:18}.


\section{Results from the models with low density wind only}
\label{results_low}

In this Section we focus on the models consisting of a pre-explosion
RSG and a low density wind only.
Most of the previous analytical and numerical studies of the
X-ray emission from SNe~II were devoted to this class of models 
\citep[see, for example,][]{chevalier:94,chevalier:03,chevalier:16,nymark:06}.

\subsection{Hydrodynamical evolution of the models and their X-ray signal}

\begin{figure}
  \centering
  \includegraphics[width=0.475\textwidth]{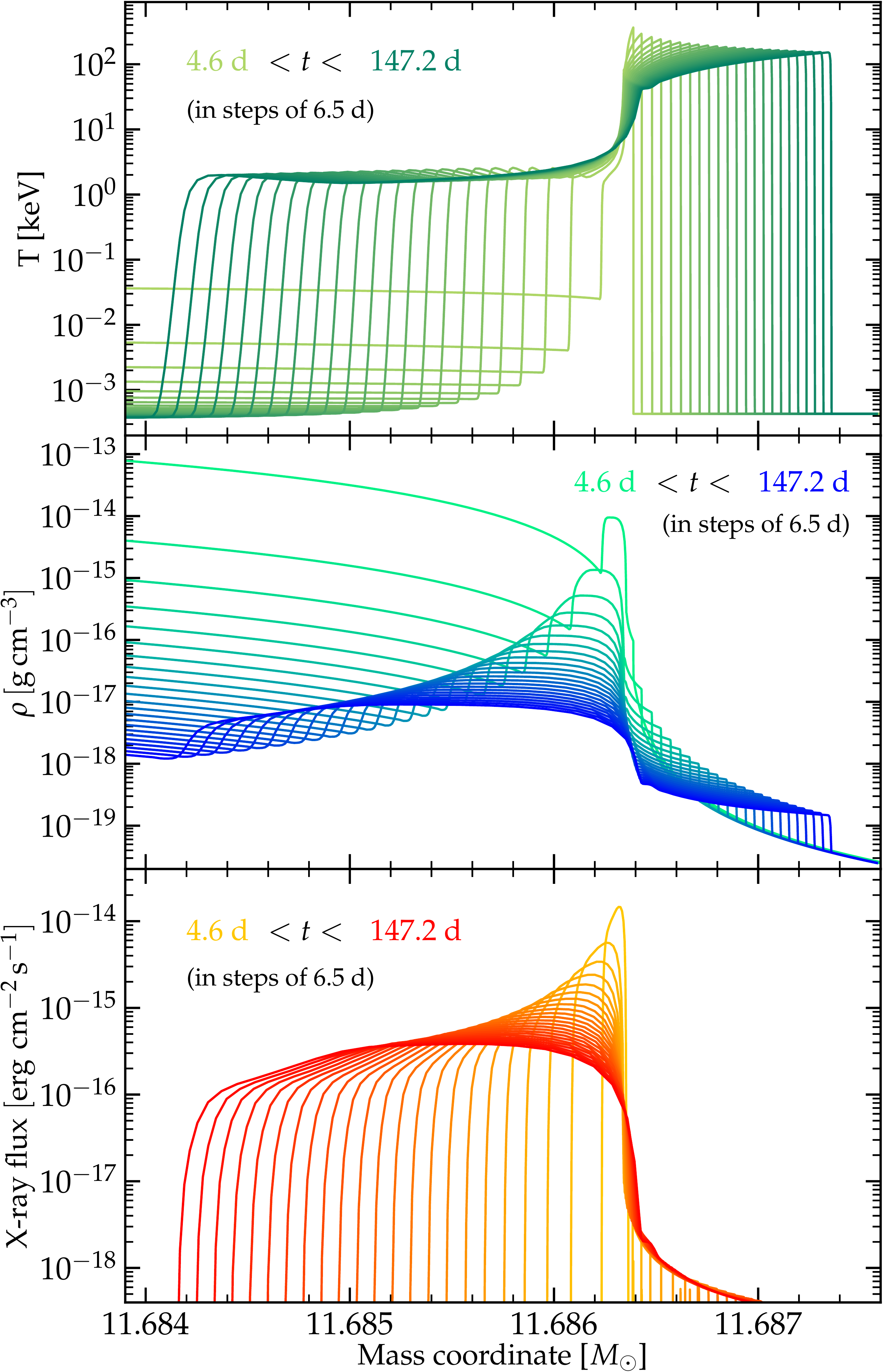}
  \caption{Evolution of the temperature profile (top panel), the density 
  profile (middle panel), and the X-ray emission from each grid
  cell (bottom panel) of the forward and reverse shock waves in
  $13\,M_{\odot}$ RSG model with $\dot{M}=2\times10^{-6}\,M_{\odot}\,{\rm yr}^{-1}$
  wind. This plot shows that the most part of X-ray emission comes from the
  dense part of ejecta between the forward and reverse shock waves. 
  The hot forward shock wave makes marginal contribution to the signal,
  justifying our assumption of equipartition between ions and electrons.
  At the same time, the X-ray emission coming from the front of the reverse shock
  wave is largely absorbed by the dense intermediate part of the shocked
  ejecta.} 
  \label{fig:hydro}
\end{figure}

Figure~\ref{fig:hydro} describes the hydrodynamical evolution of  
$13\,M_{\odot}$ RSG model with a $\dot{M}=2\times10^{-6}\,M_{\odot}\,{\rm yr}^{-1}$
wind. After the initial post-explosion shock wave
reaches the surface of the RSG, it gives rise to a forward shock wave
propagating into the low density wind and to a reverse shock wave 
propagating backwards into the shocked ejecta. The top panel of 
Figure~\ref{fig:hydro} shows the temperature profile of the two shock waves,
and the middle panel shows their density. The bottom panel of 
Figure~\ref{fig:hydro} shows the total X-ray flux from each numerical grid cell after 
taking into account the absorption by all material above that cell, 
i. e., as seen by a remote observer.

\begin{figure}
  \centering
  \includegraphics[width=0.475\textwidth]{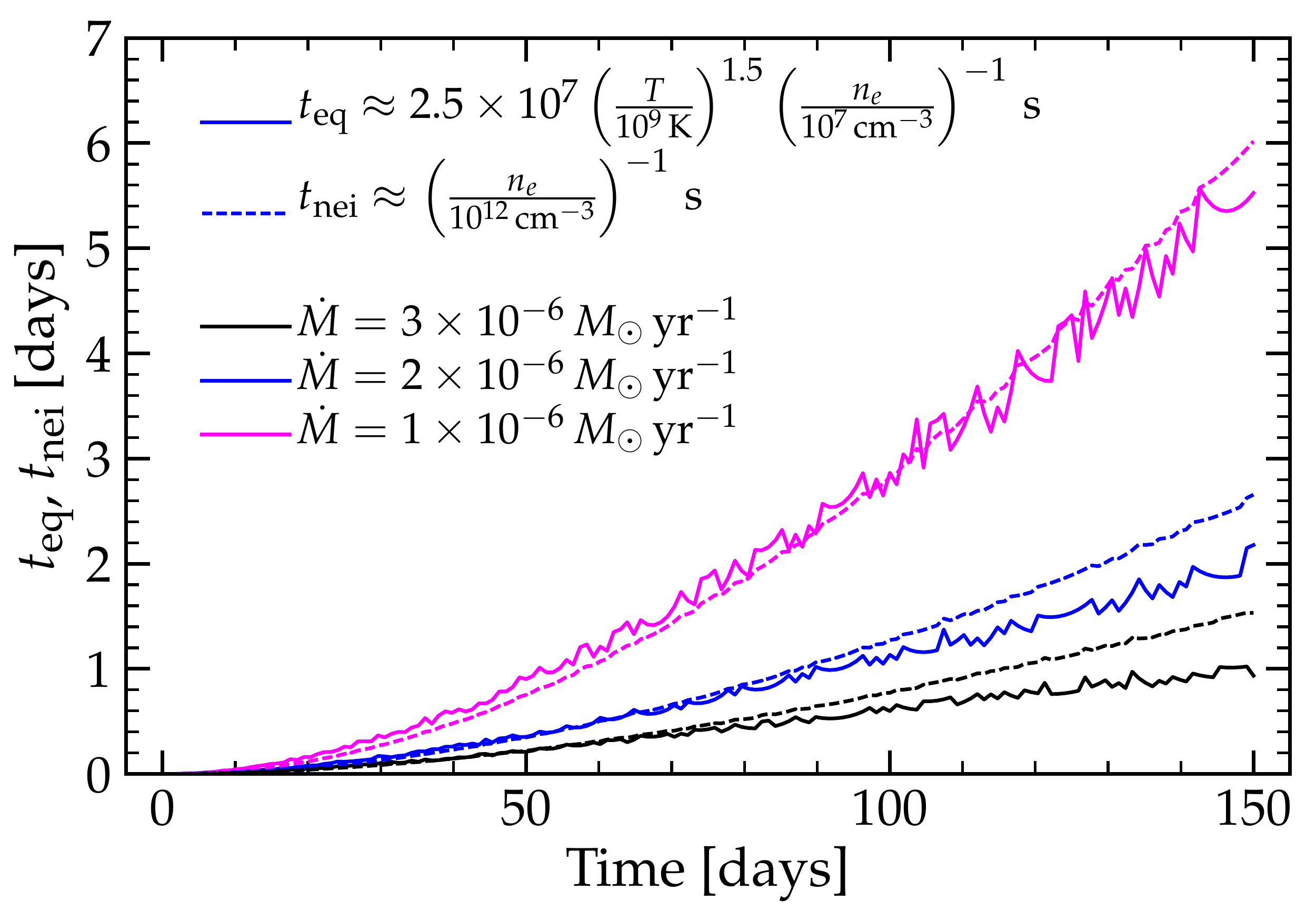}
  \caption{The equipartition timescale between ions and electrons, $t_{\rm eq}$,
  as well as the timescale $t_{\rm nei}$, which is needed to achieve equilibrium 
  ionization, calculated at the point of maximum X-ray
  flux in our models.} 
  \label{fig:timescale}
\end{figure}

Figure~\ref{fig:hydro} shows that the majority of the X-ray flux
is coming from the shocked ejecta at the interface between the 
forward and the reverse shock waves. The forward shock wave itself
generally has too high temperatures and too low densities to 
produce detectable X-ray signal, while the radiation from the front of
the reverse shock wave is largely absorbed by the external
layers. It is clear from Figure~\ref{fig:hydro} that the maxima of X-ray radiation 
closely follow the regions of maximum density, which justifies
the assumption of equality between the electron and ion temperatures
used in our analysis. While it is known that due to the slow Coulomb 
equipartition the electron temperature immediately behind the
reverse and especially the forward shock wave may be much smaller than
the ion temperature, the dense interface region between the 
shocks is expected to be in equipartition at early times \citep[see, for example,
Fig.1 of][]{chevalier:16}. To better illustrate this, in Figure~\ref{fig:timescale}
we plot the equipartition timescale, $t_{\rm eq}$, given by Eq.~\ref{equipartition}, as well as 
the timescale $t_{\rm nei}$, which is needed to achieve equilibrium ionization
\citep{dwarkadas:10}. Both timescales shown in Figure~\ref{fig:timescale} are
calculated at the point of maximum of X-ray flux in our models. By day $\sim 150$
they are still small with respect to the light curve time.

\begin{figure}
  \centering
  \includegraphics[width=0.475\textwidth]{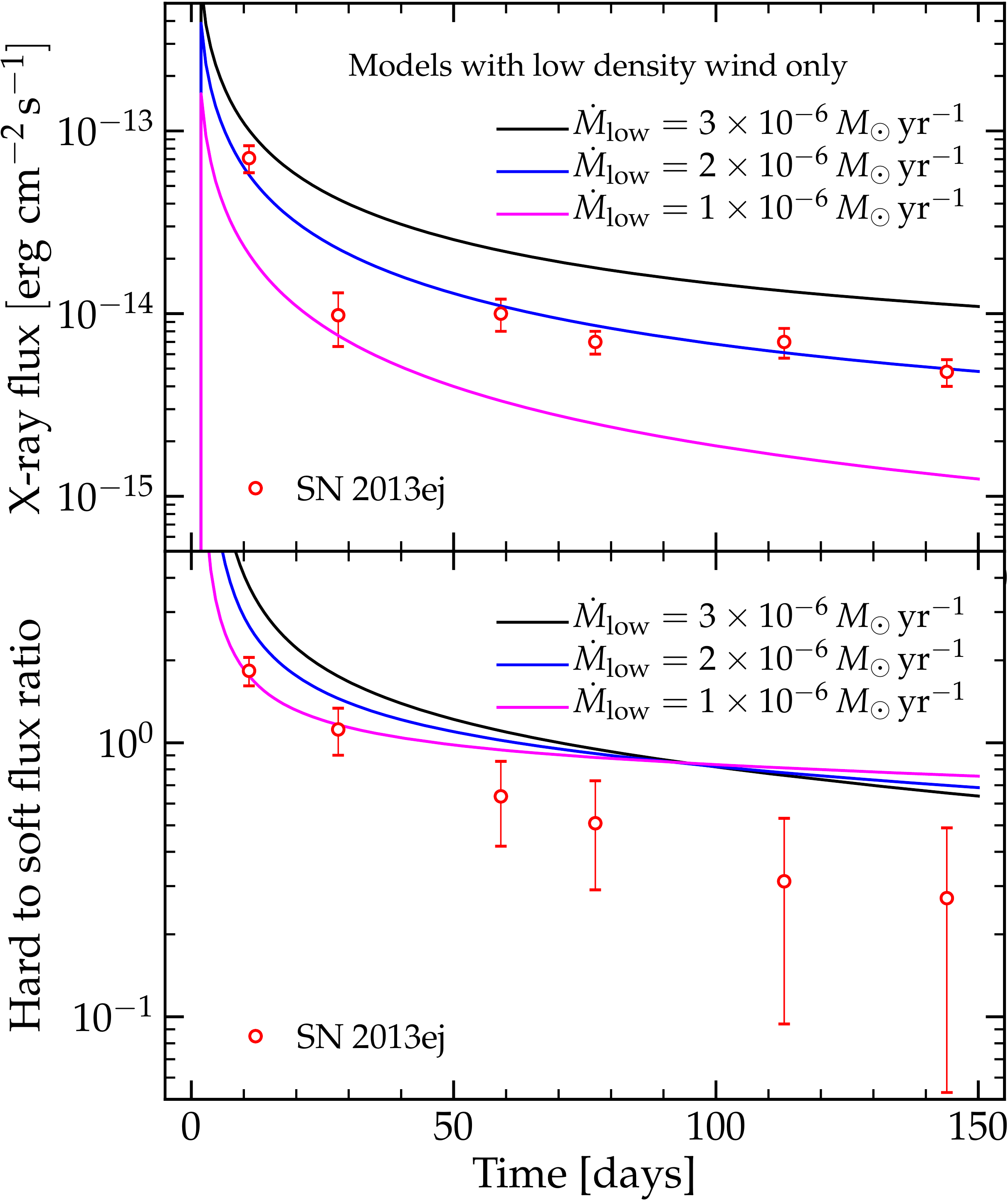}
  \caption{Top panel: Total X-ray flux obtained from 
  $13\,M_{\odot}$ RSG model with the low density wind of
  different mass loss rates $\dot{M}$, compared to the 
  observational data of SN 2013ej. Bottom panel: The ratio between hard ($2.0-8.0\,{\rm keV}$)
  and soft ($0.5-2.0\,{\rm keV}$) components of the obtained X-ray signal, compared
  to the values observed in SN 2013ej.} 
  \label{fig:winddenX}
\end{figure}

In Figure~\ref{fig:winddenX} we show the sum of the X-ray contributions
from all individual numerical cells after taking into account the external 
absorption for each cell.
The top panel of Figure~\ref{fig:winddenX} shows the total X-ray
flux we obtain from the models with low density wind only having 
the mass loss rates of $1\times10^{-6}$, 
$2\times10^{-6}$ and $3\times10^{-6}\,M_{\odot}\,{\rm yr}^{-1}$. 
The model with $\dot{M}=2\times10^{-6}\,M_{\odot}\,{\rm yr}^{-1}$
reproduces well the observed total flux from SN 2013ej.
This agrees with the result of \citet{chakraborti:16}, where the
mass loss rate from SN 2013ej was estimated as
$\dot{M}=(2.6\pm0.2)\times10^{-6}\,M_{\odot}\,{\rm yr}^{-1}$, 
based on the \texttt{XSPEC} fit of the X-ray spectra. 

At the same time, the bottom panel of Figure~\ref{fig:winddenX} 
shows that our model with $\dot{M}=2\times10^{-6}\,M_{\odot}\,{\rm yr}^{-1}$
cannot reproduce
correctly the ratio between the soft and the hard components of the
X-ray signal. In the observed data, the soft component starts to dominate
already after day $\sim30$, and by day $\sim150$ it is about 4 times
more energetic than the hard component. On the other hand, in our 
model the ratio between the soft and hard X-ray components
by the end of simulation is only $\sim 1.5$.

\subsection{Dependence on the progenitor ZAMS mass and final energy}

\begin{figure}
  \centering
  \includegraphics[width=0.475\textwidth]{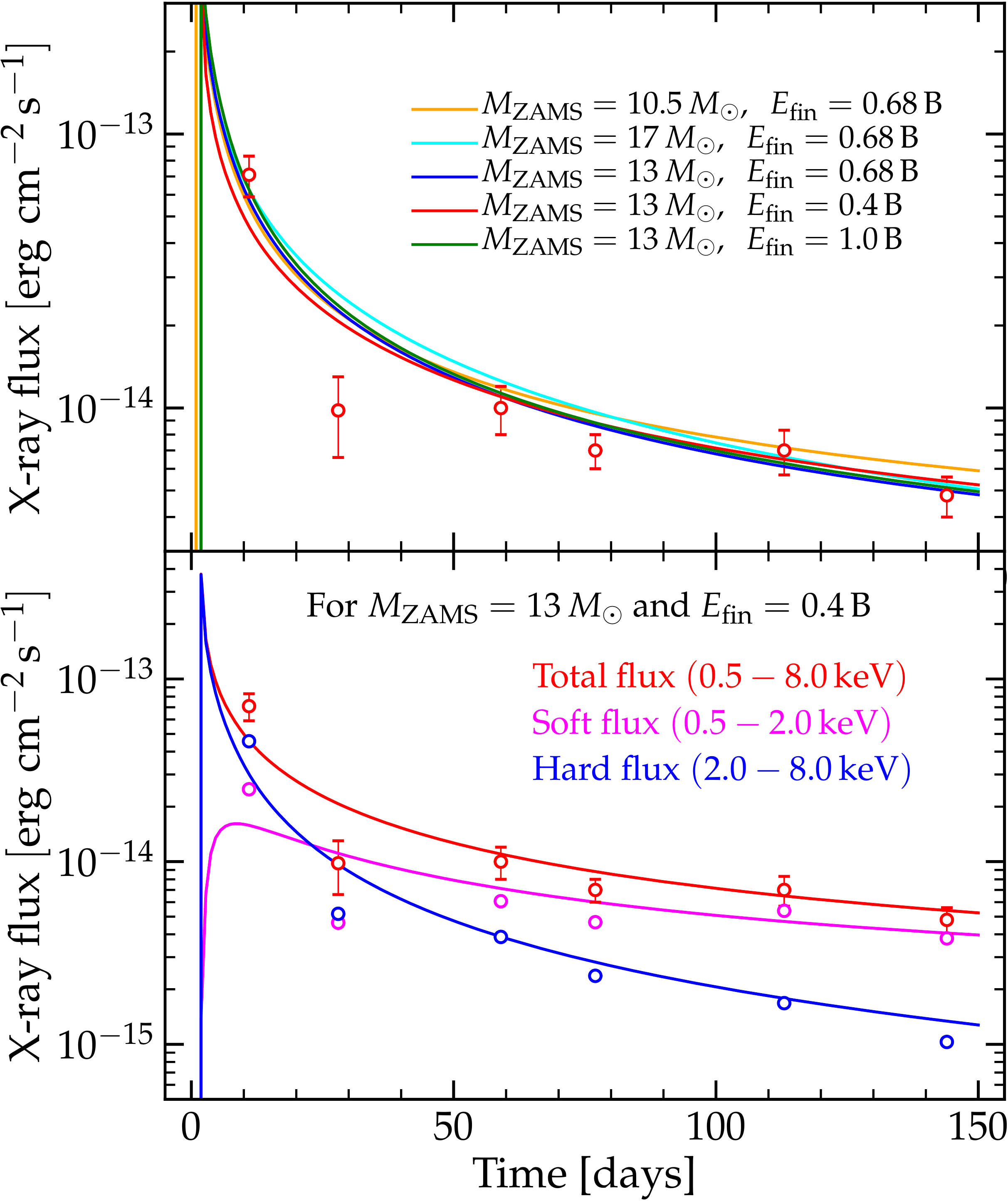}
  \caption{Top panel: Total X-ray flux from the models with different progenitor
  ZAMS mass and final energy compared to the fiducial model. Bottom
  panel: Total, soft and hard X-ray flux for the model with $M_{\rm ZAMS}=13\,M_{\odot}$
  and $E_{\rm fin}=0.4\,{\rm B}$, which demonstrates the best agreement with
  the X-ray data among the models with low density wind only.} 
  \label{fig:energy}
\end{figure}

\begin{figure}
  \centering
  \includegraphics[width=0.475\textwidth]{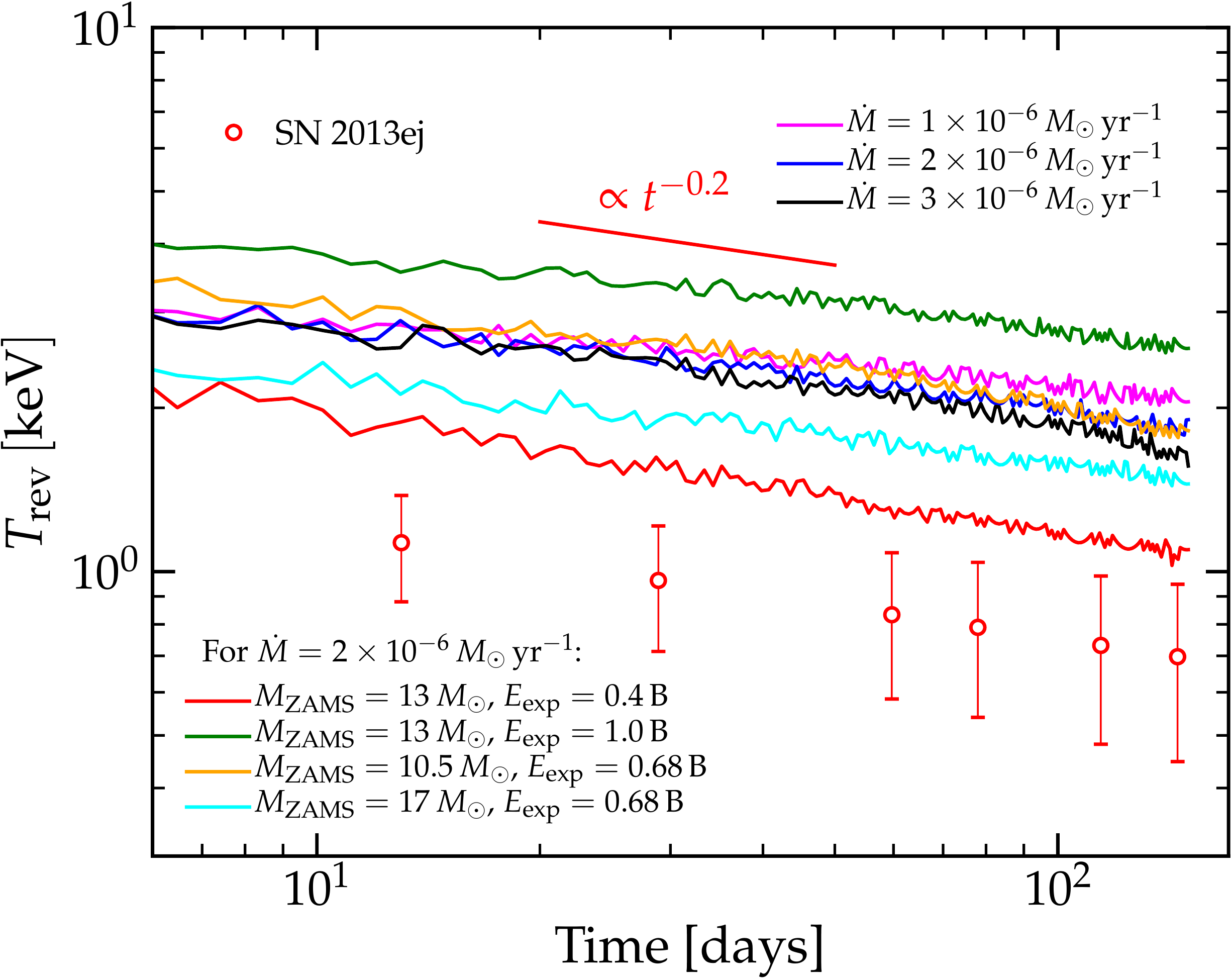}
  \caption{Comparison of the ejecta temperature at the maximum 
  of X-ray flux in our models to the reverse shock temperature derived from fitting
  the observed X-ray spectra in \citet{chakraborti:16}. The model 
  with $M_{\rm ZAMS}=13\,M_{\odot}$
  and $E_{\rm fin}=0.4\,{\rm B}$ shows the best agreement with the
  observed hard to soft flux ratio, even though its ejecta temperature is slightly
  higher than the reverse shock temperature from \citet{chakraborti:16}. All the models
  demonstrate correct $\propto t^{-0.2}$ behavior.} 
  \label{fig:templow}
\end{figure}

To check the dependence of this result on the model parameters,
we performed the simulations for 
different progenitor ZAMS masses ($10.5$ and $17\,M_{\odot}$) 
and explosion energies ($0.4$ and $1.0\,{\rm B}$) using
mass loss rate $\dot{M}=2\times10^{-6}\,M_{\odot}\,{\rm yr}^{-1}$.
The top panel of Figure~\ref{fig:energy} shows the total X-ray flux from these 
models. As expected from the analytical works 
\citep[see, for example,][]{chevalier:16}, 
the total flux is mostly determined by the mass loss rate of the wind and 
depends weakly on the explosion energy. Instead, the shock
temperature, which is proportional to the second degree of ejecta 
velocity \citep{chevalier:16}, depends almost linearly on the explosion
energy. Therefore, the model with the explosion energy $0.4\,{\rm B}$,
shown in the bottom panel of Figure~\ref{fig:energy},
demonstrates much better agreement with the observed ratio between the
soft and hard components of X-ray flux. Figure~\ref{fig:templow} shows the ejecta
temperature at the maximum of the X-ray flux for all our models with 
low density wind. For comparison, with red markers 
we show the reverse shock temperature
estimated in \citet{chakraborti:16} from fitting the X-ray spectra with 
\texttt{XSPEC}. 

\begin{figure}
  \centering
  \includegraphics[width=0.475\textwidth]{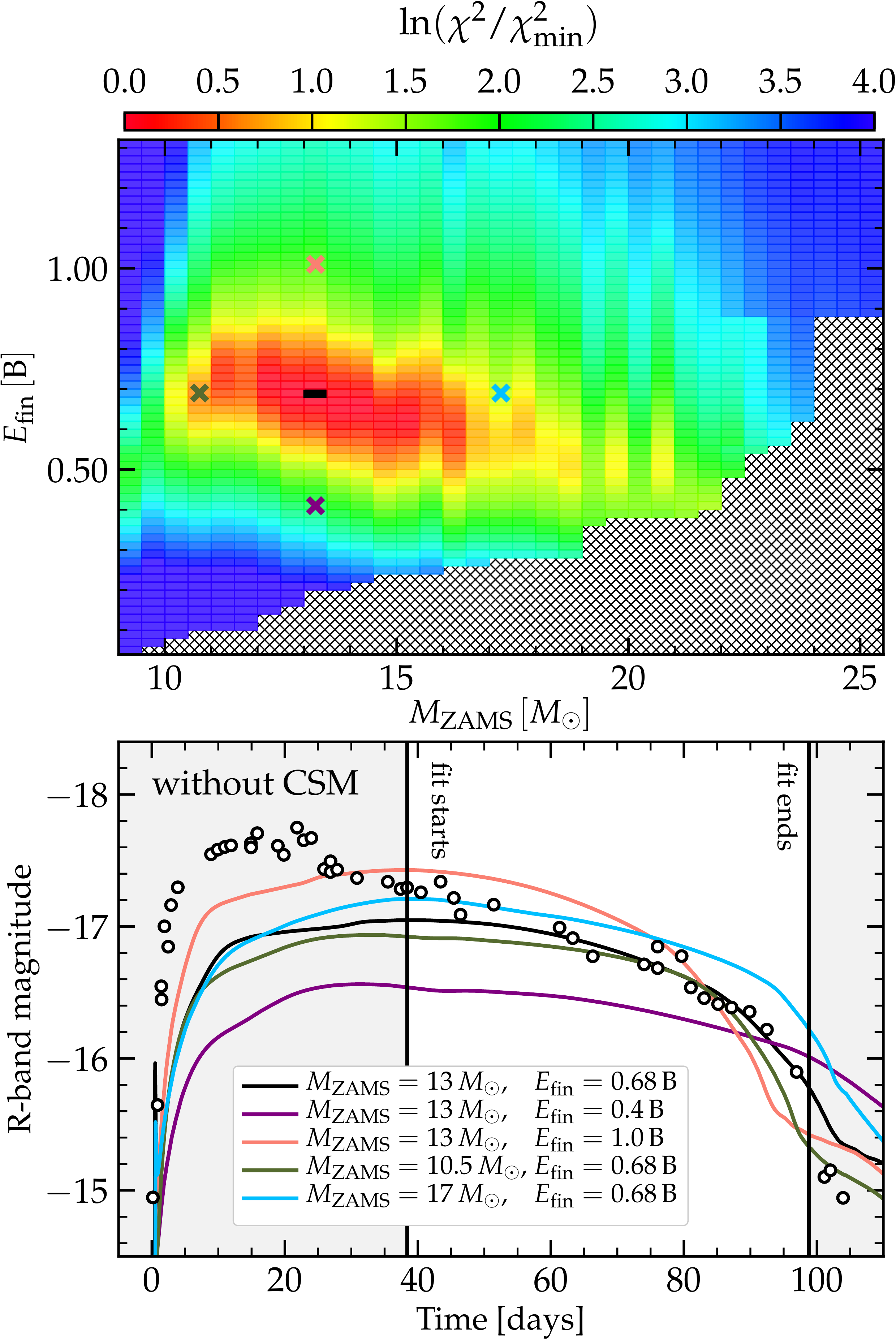}
  \caption{Top panel: Color coded ratio of $\chi^2$ to $\chi^2_{\rm min}$ at each
  point of the $M_{\rm ZAMS}-E_{\rm fin}$ grid of models without CSM for
  SN 2013ej. The black square indicates the best fitting model to the optical data between
  day 38 ($\approx$ slope break in the quasibolometric light curve of this SN) and day 99 
  (end of plateau). Bottom panel: $V$-band light curves obtained with \texttt{SNEC} 
  compared to the observational data. The best fitting model is shown in black, while the
  colored lines show the light curves from the models marked by crosses of the
  corresponding color in the top panel.} 
  \label{fig:optical}
\end{figure}

At the same time, from the numerical modeling of SN 2013ej 
optical light curve performed in 
\citet{morozova:17a}, we know that the models with progenitor ZAMS 
masses $10.5$ and $17\,M_{\odot}$ and explosion
energies $0.4$ and $1.0\,{\rm B}$, shown in Figure~\ref{fig:energy}, 
are not consistent with the optical data as well as our fiducial model. To 
illustrate this, Figure~\ref{fig:optical} shows the results obtained in \citet{morozova:17a}
for SN 2013ej without CSM. We reiterate that none of our models without a
CSM was able to reproduce the early rise and subsequent decline seen in
all (including red) bands of the optical data. Therefore, the models without CSM
were compared to the data only starting from the approximate moment of the 
slope break in quasibolometric light curve, when the impact of the CSM 
is thought to become small. This moment is marked in Figure~\ref{fig:optical} as
the start of the fit, while the end of the fit coincides with the transition from the
plateau to the radioactive $^{56}$Ni tail, when the radiation diffusion 
approach of \texttt{SNEC} is no longer valid.

The top panel of
Figure~\ref{fig:optical} shows the distribution of $\chi^2$ for the grid of models
in $M_{\rm ZAMS}-E_{\rm fin}$ parameter space from \citet{morozova:17a}, 
without CSM. Color coded is the natural logarithm of the ratio $\chi^2/\chi^2_{\rm min}$,
where $\chi^2_{\rm min}$ corresponds to the best fitting model.
The black rectangle shows the position of the best fitting model with 
$M_{\rm ZAMS}=13\,M_{\odot}$ and $E_{\rm fin}=0.68\,{\rm B}$, and the $V$-band
light curve of this model is shown in the bottom panel with black solid line.
The crosses in the top panel mark the position of the additional models, whose
X-ray signals are plotted in the top panel of Figure~\ref{fig:energy}. The light curves of these models
are shown in the bottom panel with the corresponding colors. While the 
$10.5\,M_{\odot}$ light curve is rather close to the best fit model, the other
three clearly do poorer job in representing the obesrvational data. 

\subsection{The effect of radiative cooling}
\label{sec_cooling}

To investigate the effect of radiative cooling in our 
models due to the X-ray emission, we performed
a test simulation, where it was taken into account. We
use the cooling function in the form
\begin{equation}
\label{cooling}
\Lambda = A T_6^{-\alpha} + B T_6^{\beta}\ ,
\end{equation}
where $T_6 = T/10^6\,{\rm K}$ \citep{chevalier:16}.
This function was calibrated in \citet{nymark:06} based on the hydrodynamical
simulations of the radiative shocks, including time-dependent ionization
balance and multilevel ion calculations.
We use the solar composition values of 
the parameters $A$, $B$, $\alpha$ and $\beta$ from Table 2 of 
\citet{nymark:06}, namely, 
$A=8.0\times 10^{-23}\,{\rm erg}\,{\rm cm}^3\,{\rm s}^{-1}$, 
$B=2.3\times 10^{-24}\,{\rm erg}\,{\rm cm}^3\,{\rm s}^{-1}$,  
$\alpha=0.9$ and $\beta=0.5$. We apply the cooling function 
at each time step after the initial shock wave hits the boundary between the
underlying RSG and the wind, by adding the term $-\Lambda n_e n_i dt$ 
to the energy density in the region 
between the forward and reverse shock waves, 
assuming complete ionization in this region.
The position of the forward shock is found from the temperature maximum,
while the position of the reverse shock coincides with the maximum of
the radial velocity. 

\begin{figure}
  \centering
  \includegraphics[width=0.475\textwidth]{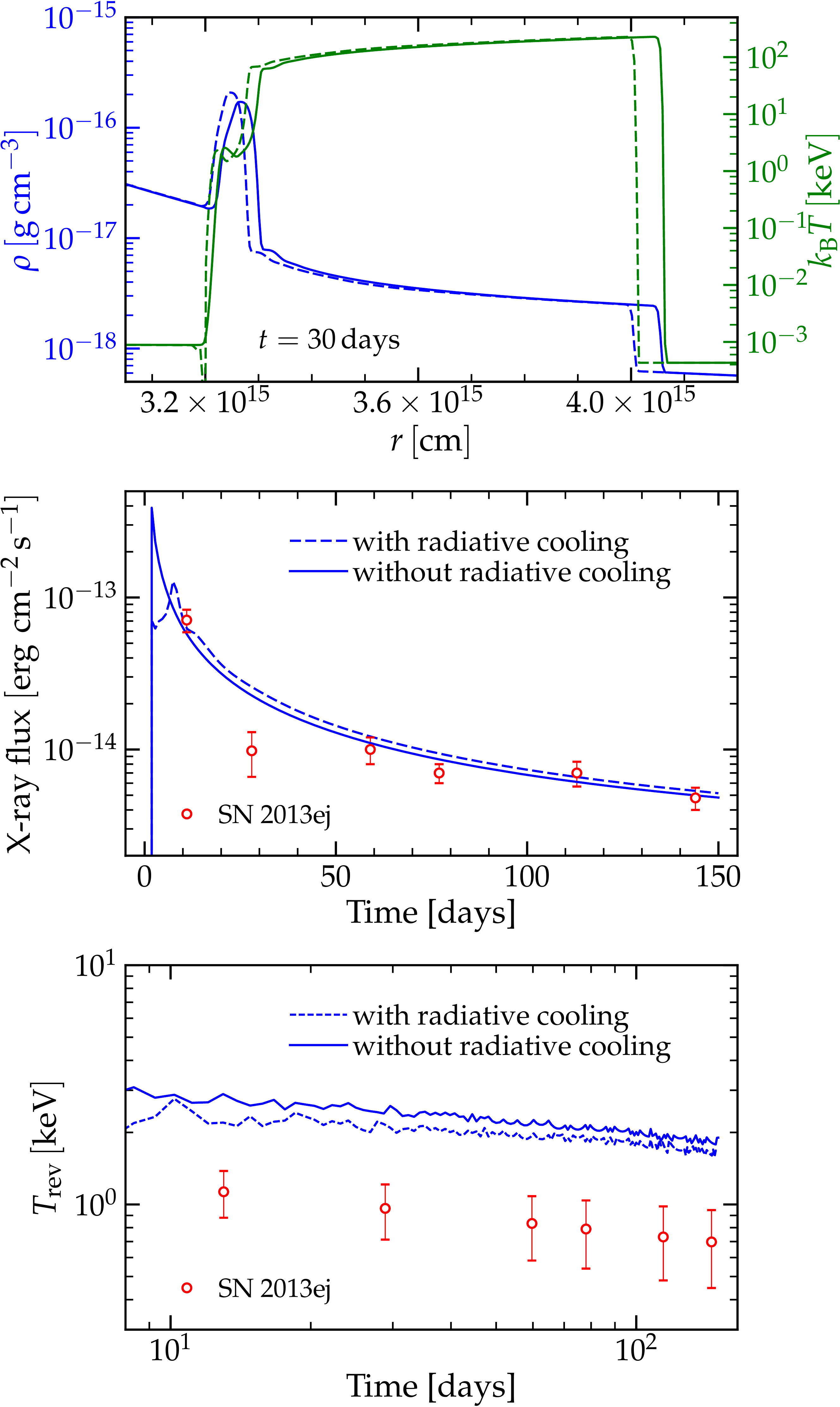}
  \caption{Top panel: Temperature and density in the shocked ejecta of
  $13\,M_{\odot}$ model with $\dot{M}=2\times 10^{-6}\,M_{\odot}\,{\rm yr}^{-1}$
  at $t=30\,{\rm days}$. Solid lines
  represent the fiducial model, and dashed lines represent
  the model taking into account radiative cooling as described in the text.
  Middle panel: Total X-ray flux from the models with and without radiative
  cooling. Bottom panel: Ejecta temperatures at the maximum of X-ray flux for the
  models with and without radiative cooling.} 
  \label{fig:cool}
\end{figure}

Figure~\ref{fig:cool} summarizes the result of this test. We find that the
radiative cooling indeed leads to a slightly higher density in the shell between the
forward and reverse shocks, as well as to a slightly lower temperature in that 
region. However, this effect is weak in our models, and does not lead to
a better agreement with the observed X-ray data. Note
that we use the cooling function calibrated for the solar composition, while the
metallicity at the explosion cite of SN 2013ej is significantly
subsolar \citep{chakraborti:16}. If anything, we expect that at lower metallicities the 
X-ray radiative cooling would be weaker. In fact, at the temperatures of interest
$\lesssim 2.24\,{\rm keV}$ ($2.6\times 10^7\,{\rm K}$) the X-ray emission 
(and cooling) is dominated by the line emission from metals \citep{chevalier:16}. 
Indeed, the metal rich compositions studied in \citet{nymark:06} resulted in
much higher values of the coefficients of Equation~(\ref{cooling}). We emphasize
that in order to obtain the X-ray signal with \texttt{XSPEC} we used the observed
subsolar value of the metallicity, $Z=0.295\,Z_{\odot}$.

To conclude this section, among the models including low density wind
only we could not find the one that would give a simultaneous fit to
the X-ray signal, including its total flux and ratio between the soft and hard
components, as well as to the optical data.

%


\section{Results from the models including dense CSM with constant slope}
\label{results_CSM}

In this section, we describe the optical and X-ray light curves from the models
consisting of a RSG, a dense CSM with the constant slope $n$, and a low density
wind.

The most commonly studied case of such CSM is the one with $n=2$,
because it has clear physical interpretation as a constant velocity steady wind.
In \citet{morozova:17}, we focused on this sort of CSM in order to explain the
early photometry of three SNe~II with various decline rates of the light curve, 
from almost flat Type IIP (plateau) SN 2013fs to the rapidly declining 
Type IIL (linear) SN 2013by, where SN 2013ej
represented the object of somewhat transitional type between the two. In the
subsequent work, \citet{morozova:17a}, we generalized this approach by
fitting the optical light curves of 20
well observed SNe~II and deducing their model parameters, again including
SN 2013ej in the set. According to our findings, the best fitting model for
SN 2013ej had $0.49\,M_{\odot}$ (here, we round up this value to $0.5\,M_{\odot}$) 
of the dense $n=2$ CSM extending up to the
radial coordinate $R_{\rm CSM}=1800\,R_{\odot}$. Therefore, our first goal
was to check whether the same progenitor model can reproduce the
X-ray light curve of SN 2013ej, after we supplement it with a low density regular RSG
wind.

\begin{figure}
  \centering
  \includegraphics[width=0.475\textwidth]{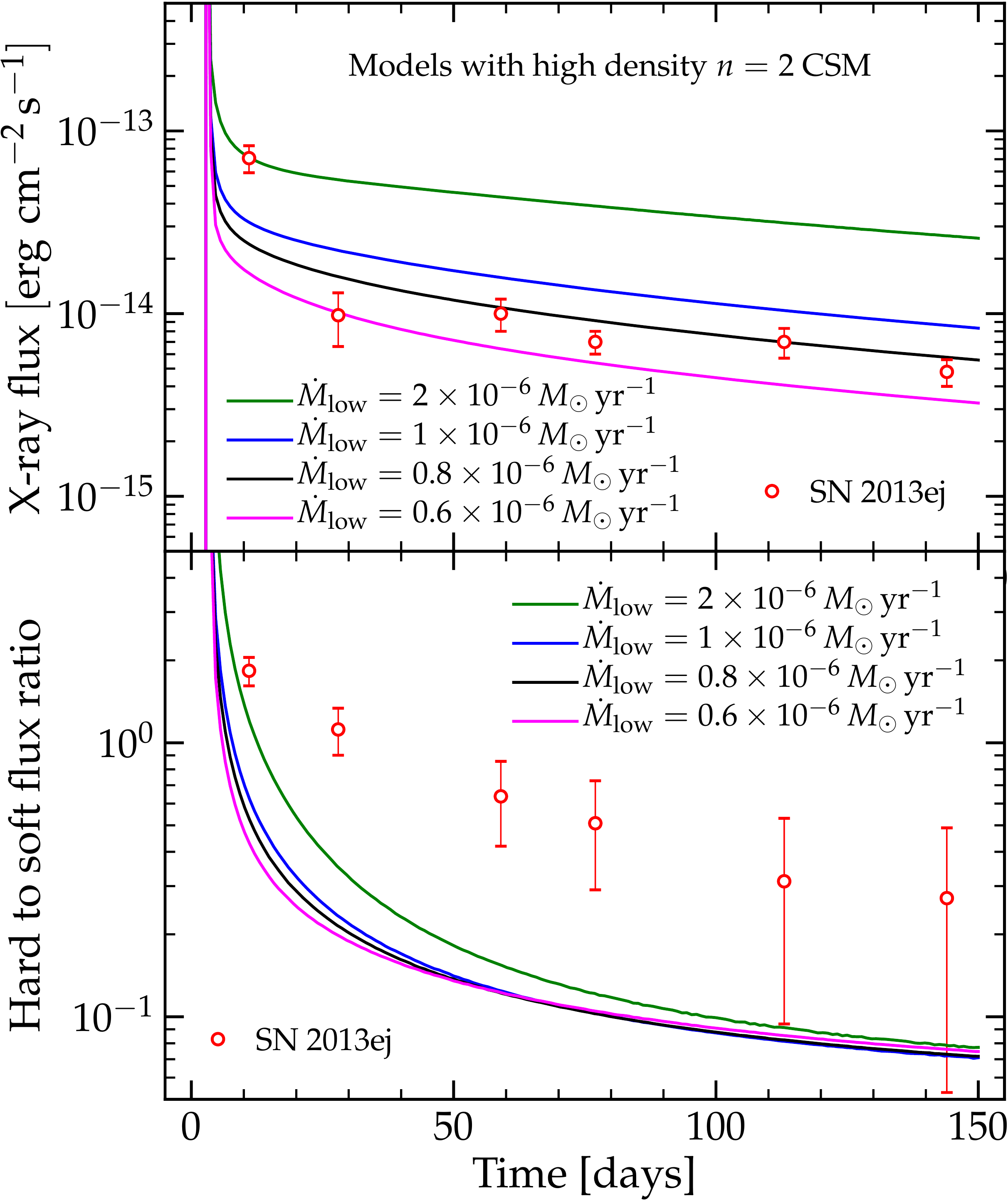}
  \caption{Top panel: Total ($0.5-8.0\,{\rm keV}$) X-ray flux from the model 
  consisting of an $M_{\rm ZAMS}=13\,M_{\odot}$ RSG, the high density steady
  wind with a total mass of $0.5\,M_{\odot}$, and the low density wind with
  different values of mass loss $\dot{M}_{\rm low}$. The final energy of the model
  is $0.68\,{\rm B}$ in all cases. Bottom panel: The ratio between hard ($2.0-8.0\,{\rm keV}$)
  and soft ($0.5-2.0\,{\rm keV}$) components of the obtained X-ray signal, compared
  to the values observed in SN 2013ej.} 
  \label{fig:windden}
\end{figure}

Figure~\ref{fig:windden} shows the X-ray light curves obtained from this model 
for different values of the mass loss in the low density wind. The final energy of
the model is $0.68\,{\rm B}$ in all cases \citep{morozova:17a}.
From the top panel of Figure~\ref{fig:windden}, one can see that the model
with $\dot{M}_{\rm low}=2\times10^{-6}\,M_{\odot}\,{\rm yr}^{-1}$ agrees well
with the total observed X-ray flux of SN 2013ej.
However, the bottom panel of this figure shows that for all values of $\dot{M}_{\rm low}$
the simulated X-ray emission is too soft compared to the observed one.

\begin{figure}
  \centering
  \includegraphics[width=0.475\textwidth]{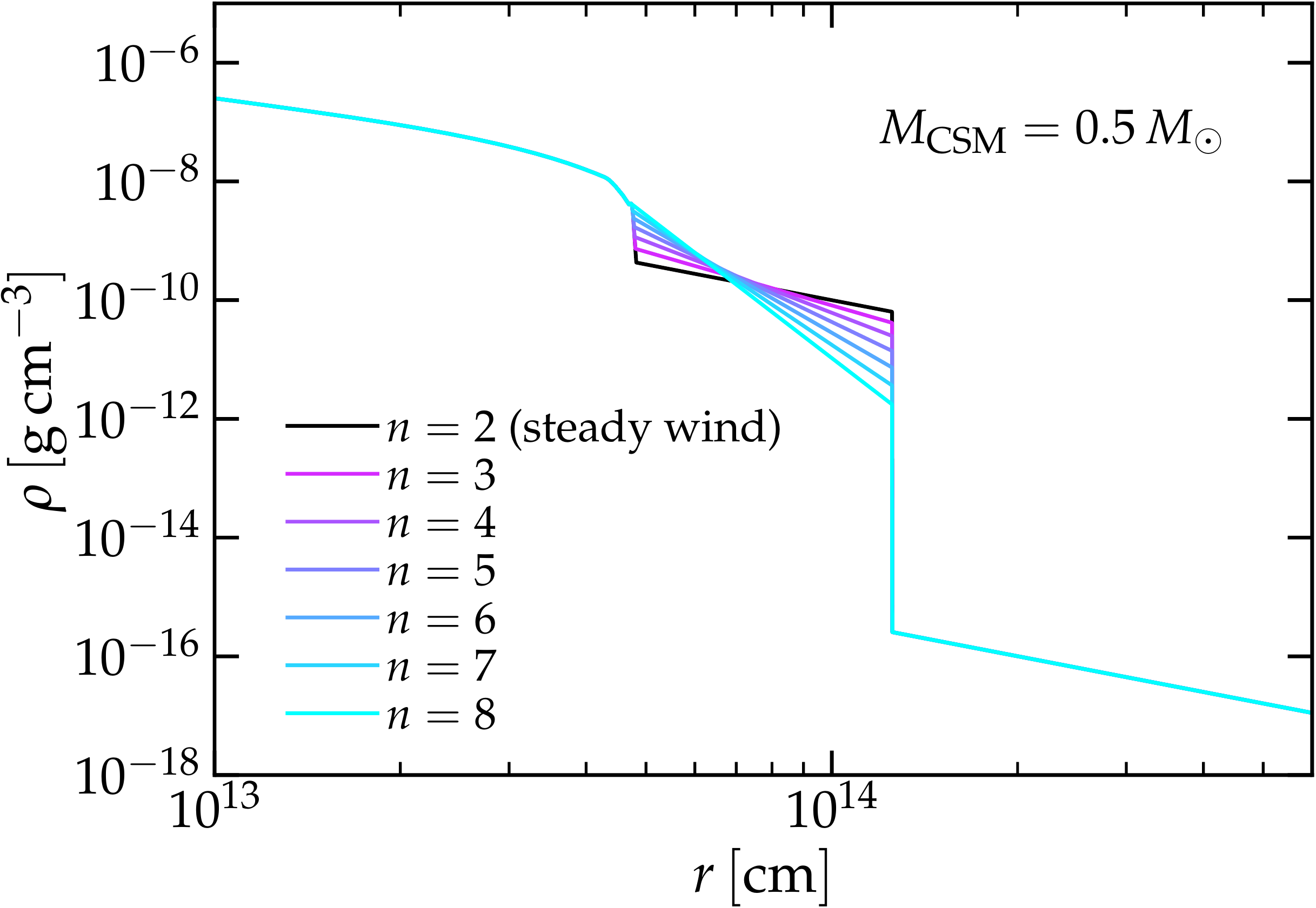}
  \caption{Density profiles of the models including dense 
  CSM with different slopes $n$.} 
  \label{fig:denacc}
\end{figure}

In this model, as well as all other models of this section, the majority of the
X-ray flux is coming from the shocked ejecta between the forward and reverse
shock waves formed at the interface between
the high density CSM and the low density wind (think of Figure~\ref{fig:hydro} at the outer
edge of dense CSM instead of the RSG surface). The softness of the synthetic 
X-ray light curve seen in the bottom panel of Figure~\ref{fig:windden} tells us that the 
temperature of the reverse shock wave is too low. One way to increase the shock 
temperature without changing the final energy of the model is to increase the slope of the CSM.
In a steeper CSM, the original post-explosion shock wave can accelerate to larger
velocites and, consequently, heat up the material to higher temperatures. Indeed,
analytical results of \citet{chevalier:16} suggest that the
temperature in the reverse shock is proportional to the squared velocity of
ejecta  (although, the slope exponent $n$ itself enters the equation for the
temperature as $(n-2)^{-2}$).

\begin{figure}
  \centering
  \includegraphics[width=0.475\textwidth]{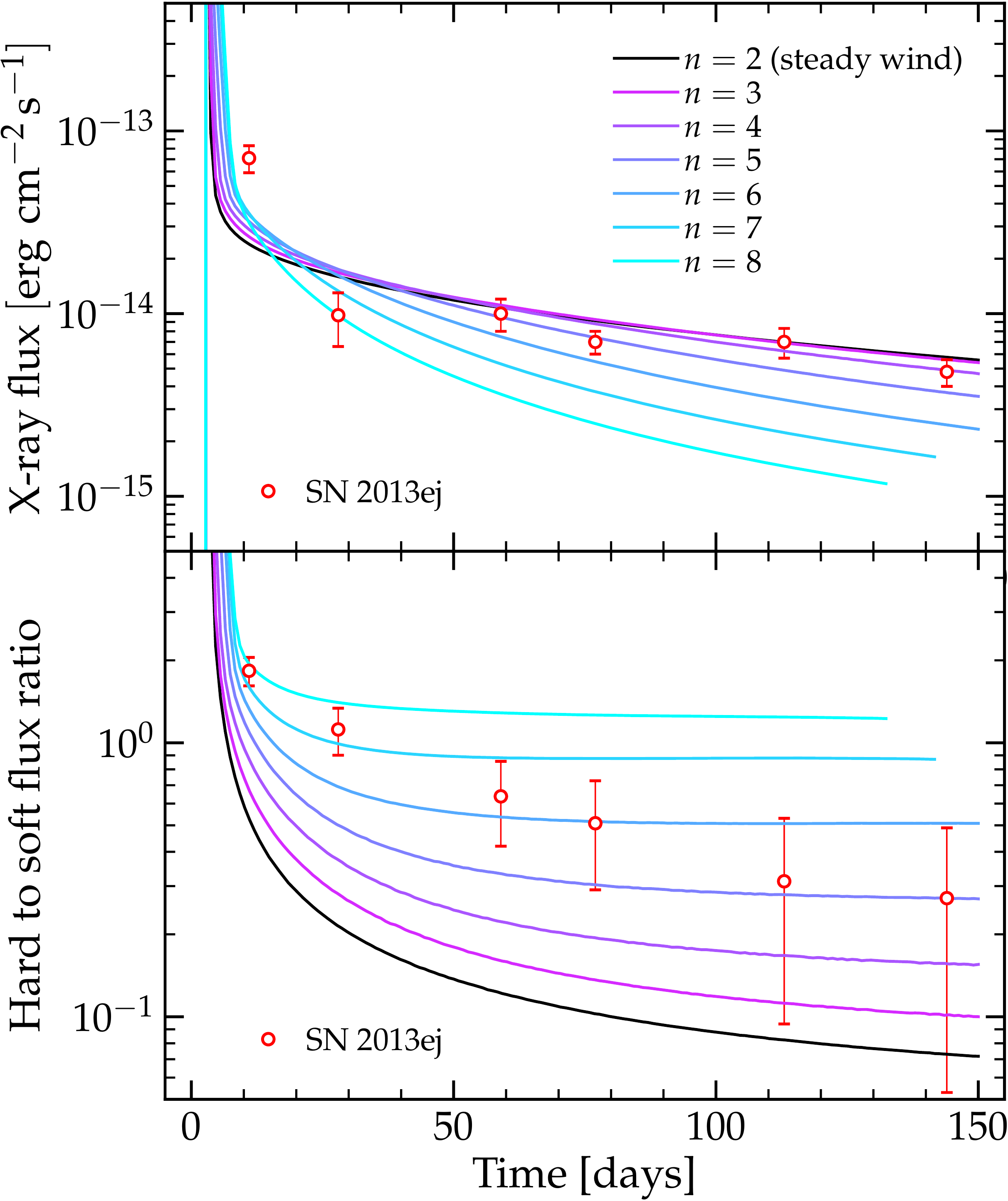}
  \caption{Top panel: Total ($0.5-8.0\,{\rm keV}$) X-ray flux from the models shown in 
  Figure~\ref{fig:denacc}. The final energy of the models
  is $0.68\,{\rm B}$ in all cases. Bottom panel: The ratio between hard ($2.0-8.0\,{\rm keV}$)
  and soft ($0.5-2.0\,{\rm keV}$) components of the obtained X-ray signal, compared
  to the values observed in SN 2013ej.} 
  \label{fig:windacc}
\end{figure}

This motivates us to consider the CSM of different slopes from 
$n=2$ to $n=8$ with the density profiles shown in 
Figure~\ref{fig:denacc} (the black curve in this plot corresponds to the steady wind
used in Figure~\ref{fig:windden}). In \citet{morozova:17a}, we demonstrated
that the confidence regions around the best fitting models in the parameter
space of the dense CSM were aligned with the isocontours of its total mass.
This led us to a conclusion that the total mass of the CSM is
better constrained by our models than its density and extent taken separately.
Therefore, when constructing the models of dense CSM in 
Figure~\ref{fig:denacc}, we imposed the condition of constant total CSM
mass, $M_{\rm CSM}=0.5\,M_{\odot}$, as derived in \citet{morozova:17a}
for SN 2013ej. 
In addition, we fixed the external radius of the CSM to $1800\,R_{\odot}$.
This agrees with the compactness of the dense CSM found in observations
\citep[see, for example,][]{yaron:17}.

\begin{figure}
  \centering
  \includegraphics[width=0.455\textwidth]{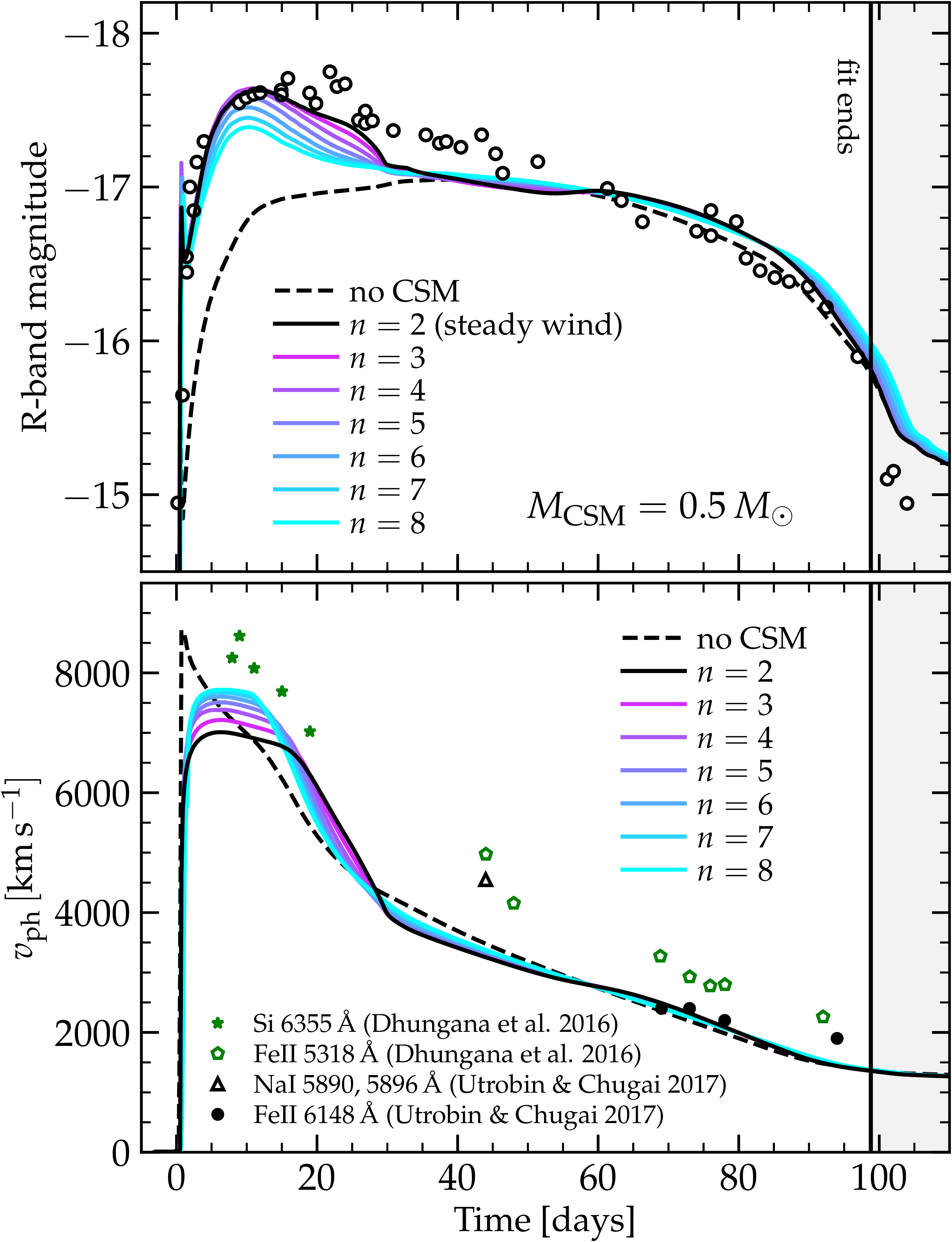}
  \caption{Top panel: $R$-band optical light curves obtained from the models shown in 
  Figure~\ref{fig:denacc}. For comparison, black dashed line shows the light curve
  from a bare RSG without the dense CSM. The gray
  shaded area marks the region after the end of plateau, 
  where the radiation diffusion approach used in \texttt{SNEC} is not
  valid. The total CSM mass of $0.5\,M_{\odot}$ is common between the models
  with different slope $n$. Bottom panel: Photospheric velocities
  of the same models versus the line velocities measured from the spectra of SN 2013ej
  in \citet{dhungana:16} and \citet{utrobin:17}.} 
  \label{fig:degrees}
\end{figure}
\begin{figure}
  \centering
  \includegraphics[width=0.41\textwidth]{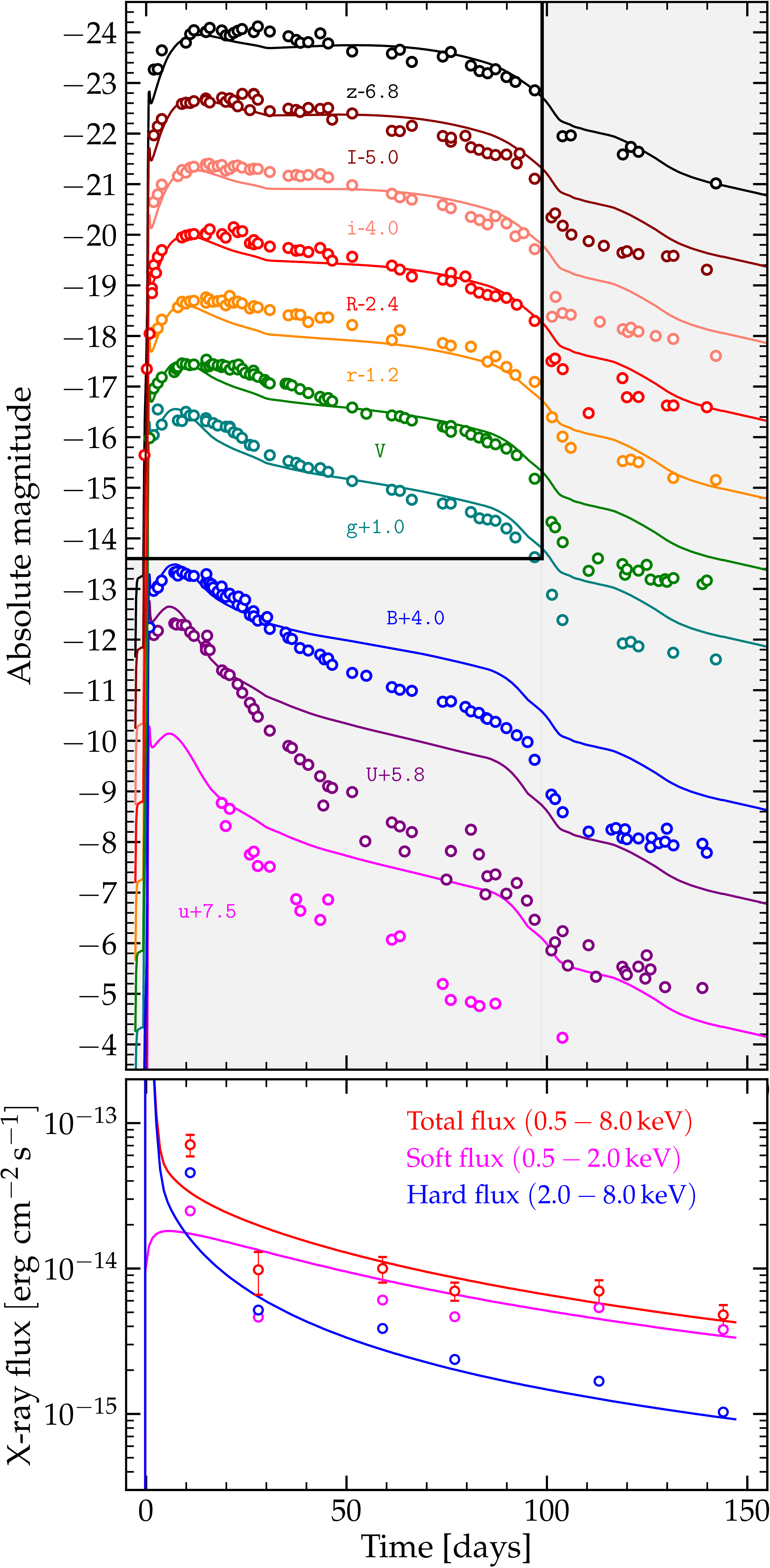}
  \caption{The model from our set, which demonstrates agreement with both optical
  (top panel) and X-ray (bottom panel) data of SN 2013ej. The parameters of this model
  are: $M_{\rm ZAMS}=13\,M_{\odot}$, $E_{\rm fin}=0.68\,{\rm B}$, 
  $M_{\rm CSM}=0.5\,M_{\odot}$, $R_{\rm ext}=1800\,R_{\odot}$, $n=5$,
  $\dot{M}_{\rm low}=0.9\times10^{-6}\,M_{\odot}\,{\rm yr}^{-1}$.} 
  \label{fig:fit}
\end{figure}

In Figure~\ref{fig:windacc}, we plot the total X-ray flux (top panel) and the 
ratio between the hard and soft components of this flux (bottom panel) from the models 
shown in Figure~\ref{fig:denacc}. It confirms our expectation that the X-ray signal 
becomes harder with the increasing slope of the CSM. The total X-ray flux changes 
as well, but to a smaller extent. 

In the top panel of Figure~\ref{fig:degrees}, 
we plot the optical ($R$-band) light curves obtained from
the same models using \texttt{SNEC}. Black markers show the data observed
from SN 2013ej. Gray shaded region
marks the end of plateau, after which the assumptions used in \texttt{SNEC} are
no longer valid. For comparison, the model without the dense CSM is shown by
the black dashed line (the same as black solid line in the bottom panel of 
Figure~\ref{fig:optical}).
Figure~\ref{fig:degrees} shows that the CSM profiles with $n$
between 2 and $\sim$5 produce very similar light curves. 
The light curves start to deviate significantly only
for the steep CSM profiles with $n>5$. However, taking into account that the 
difference in density between the $n=2$ and $n=8$ CSM profiles reaches few orders 
of magnitude (see Figure~\ref{fig:denacc}), we find that the maximum variation of
$\sim$$0.4\,{\rm mag}$ between their light curves is rather small. This supports our
assumption that the total CSM mass is one of the main parameters constrained
by the hydrodynamical models of optical light curves.

The bottom panel of Figure~\ref{fig:degrees} shows the velocity
at the photosphere location, $v_{\rm ph}$, of the models from Figure~\ref{fig:denacc}. 
In the literature,
the photospheric velocity is commonly compared to the velocities of Si or Fe lines 
measured from the 
SN spectra. In the bottom panel of Figure~\ref{fig:degrees}, the green markers indicate
the velocities of Si $6355\,${\AA} and FeII $5318\,${\AA} lines found 
from the spectra of SN 2013ej in \citet{dhungana:16}. There are many more
lines analyzed in the work of \citet{dhungana:16}, and
the velocities of different lines measured from the same spectra show very large scatter, 
differing by $2000-4000\,{\rm km}\,{\rm s}^{-1}$ and more (see their Figure 20), but the
Si $6355\,${\AA} and FeII $5318\,${\AA} lines correspond to the lowest velocities found 
at the early and late times, respectively. In addition, the black markers 
in the bottom panel of Figure~\ref{fig:degrees} indicate the
velocities of NaI $5890$, $5896\,${\AA} doublet and FeII $6148\,${\AA} line found 
from the same spectra by \citet{utrobin:17}. The largest disagreement between the
measured line velocities and the photospheric velocity of our models is seen on day 44.
In general, the qualitative agreement between the photospheric and line velocities
that we find in Figure~\ref{fig:degrees}
is similar to the one seen in the models of \citet{utrobin:17}, which may be
surprising given the large difference between the best fitting model parameters
derived for SN 2013ej in \citet{morozova:17a} and \citet{utrobin:17}. 

However, we would
like to emphasize that it is not at all obvious that the photospheric velocity of the numerical
models should coincide with the Fe or Si line velocities measured from the spectra.
Instead, the recent work of \citet{paxton:18} demonstrated that the material velocity at the location 
where the Sobolev optical depth of the FeII $5169\,${\AA} line, $\tau_{\rm Sob}$, is equal to $1$
provides a much better match to the observations than the photospheric velocity. 
They find that the velocities found at $\tau_{\rm Sob}=1$ and $\tau=2/3$ start to
disagree substantially after day $\sim$30 and may differ by up to a $1000\,{\rm km}\,{\rm s}^{-1}$
and more. Currently, we do not have the capability to estimate the $\tau_{\rm Sob}=1$
velocity of the FeII $5169\,${\AA} line from the hydrodynamical output of \texttt{SNEC}, but
we plan to work on it in the future.

Finally, Figure~\ref{fig:fit} represents the main result of this section by showing a 
candidate model that can reproduce both optical and X-ray data of SN 2013ej. 
Here, we summarize the parameters of this model. It is
based on the $13\,M_{\odot}$ (at ZAMS) RSG having final energy of $0.68\,{\rm B}$,
which is necessary to reproduce the plateau part of the observed optical data.
It has $0.5\,M_{\odot}$ of the dense compact ($R_{\rm CSM}=1800\,R_{\odot}$) CSM,
the presence of which is suggested by the rapidly rising early optical data. The 
density in the CSM drops as $\rho\propto r^{-5}$, which ensures that its
outermost region is sufficiently accelerated in order to reproduce the ratio between
the hard and soft components of the observed X-ray emission. Above the dense
CSM, this model has regular low density RSG wind with mass losses 
of $\dot{M}_{\rm low}=0.9\times10^{-6}\,M_{\odot}\,{\rm yr}^{-1}$ needed to 
reproduce the total X-ray flux from SN 2013ej.


Since our numerical simulations involve many approximations in both optical
and X-ray components (see Sections~\ref{athena}-\ref{snec}), the model shown 
in Figure~\ref{fig:fit} cannot be taken as a final word on the progenitor of SN 2013ej. 
Moreover, in this study we did not attempt
to minimize the residuals and find an optimal fit across a grid of all parameters
used in our models. Nevertheless, there are few general conclusions that
can be drawn from this section. First of all, it shows that inclusion of a dense CSM
does not make the model incompatible with the observed X-ray emission.
In other words, the qualitative 
agreement with the early X-ray signal seen in the bottom panel of
Figure~\ref{fig:fit} is as good as the one from Figure~\ref{fig:energy}.
The main difference between these two scenarios is in the location of the
reverse shock wave, which produces the bulk of the X-ray emission. In
the models with low density wind only, this reverse shock wave probes the surface
region of the RSG, while in the models with the dense CSM, it probes the outermost
layer of the CSM.

Second, this result shows that the early X-ray signal may be used as an important
complementary piece of information in constraining the properties of the dense
CSM surrounding RSGs before explosion. Analysis of this section shows
that the CSM with constant $n=2$ slope along its entire extension produces
the reverse shock wave that is too cold to fit the observed X-ray emission
in the first few tens to hundreds of days. In reality, it is likely that the radiative cooling during
and after the shock breakout will make the temperature of the shock wave 
even lower, and the X-ray emission even softer for the same CSM configuration.
Therefore, we conclude that the density profile should be steeper than $n=2$ at 
least in some regions of the dense CSM, if not along its entire extension,
to ensure the efficient acceleration of its outermost layer. Generally, this conclusion is
in agreement with the idea of accelerating wind, 
proposed in the works of \citet{moriya:17,moriya:18}.
At the same time, for now we refrain from explaining the origin of the dense
CSM by either accelerating or steady wind, because our models are not sensitive
to the pre-explosion velocities in the CSM (as long as they are much smaller than the
post-explosion velocities of a few tens of thousands of ${\rm km}\,{\rm s}^{-1}$) 
and, therefore, cannot constrain its formation history. We consider the models
even more similar to the ones from \citet{moriya:17,moriya:18} in the following section.

In regard to the formation mechanism of the CSM, one thing that
we would like to emphasize is its compactness. The difference between the external 
radius of our best fitting model and the pre-explosion radius of the 
$M_{\rm ZAMS}=13\,M_{\odot}$ \texttt{KEPLER} progenitor is $\sim$1100$\,R_{\odot}$.
Assuming that is expands with the velocity of $100\,{\rm km}\,{\rm s}^{-1}$, 
the formation of such CSM 
would take no longer than $\sim$3 months prior to the explosion. This value of the velocity was 
quoted as an upper limit and used in the analysis of the early spectra of SN 2013fs
by \citet{yaron:17}. At the same time, they mention that Doppler broadening of the
O \texttt{VI} doublet observed in the early spectra of SN 2013fs may be compatible with
even higher velocities in the inner parts of the CSM. Recent analysis of the pre-explosion
images of four SNe~II (including SN 2013ej, the object of our study) by \citet{johnson:17}, 
demonstrated that they show no signs
of variability in the last several years before the explosion. However, the 
number of the pre-explosion images obtained in the last few months prior to the
SN explosions is still very small. Therefore, we believe that there
is not yet enough observational evidence to rule out entirely the possibility of an eruptive outburst
in the last months of the RSG evolution\footnote{One exception is SN 2013am, for which
the last pre-explosion observation was taken mere 5 days before the explosion,
with no signs of variability \citep{johnson:17} . However,
there are noticeable qualitative differences in the behavior of light curves of SNe 2013am and
2013ej. For example, the $I$-band light curve of SN 2013am shows a very long steady
rise lasting more than $\sim$50 days before it reaches the maximum and starts to decline
(see \citealt{zhang:14} and \url{https://sne.space/} by \citealt{guillochon:17}).
At the same time, the $I$-band light curve of SN 2013ej reaches its peak in 
less than 30 days and shows a steady decline after that. The fast early rise and decline
observed in the red bands of some SNe is one of the most difficult things to reproduce 
by the theoretical light curves of bare RSGs, and it was one of the reasons to include the 
CSM in the models of \citet{morozova:17,morozova:17a}. Though this is not in the focus
of our study, we find it very likely that the light curve of SN 2013am could be explained by
the explosion of a bare RSG without the dense CSM.}.

\section{Results from the models including dense CSM with varying slope}
\label{results_Mor}

In this section, we present the X-ray light curves obtained from the models with
dense CSM of a variable slope, constructed in a manner similar to the one from
the works of \citet{moriya:17,moriya:18}. We could not generate the optical light
curves of these models with the curent version of \texttt{SNEC} due to the
low densities in the CSM. Nevertheless, based on the results of \citet{moriya:17,moriya:18}, 
we expect that these models could provide a reasonable fit to the optical data.

\begin{figure}
  \centering
  \includegraphics[width=0.475\textwidth]{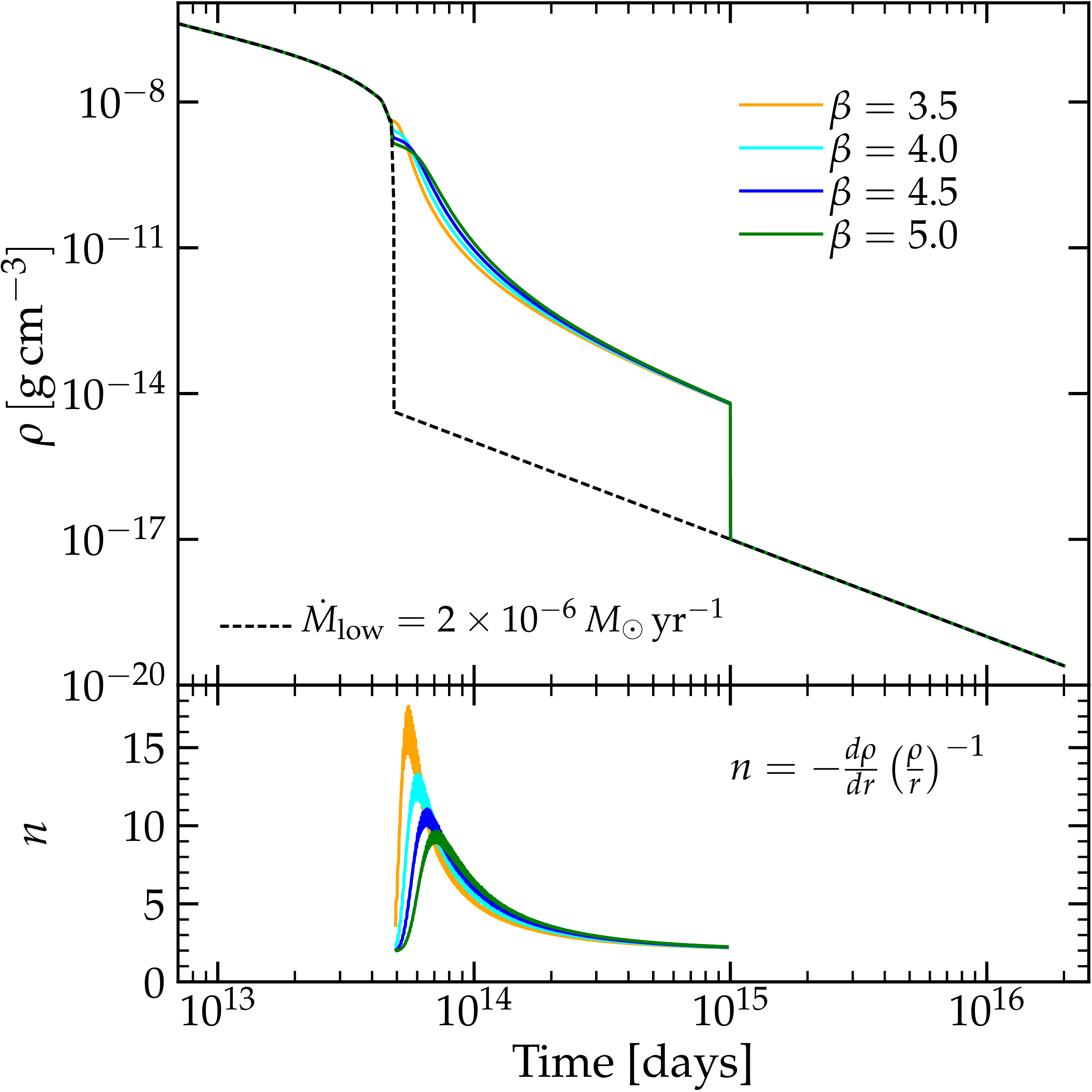}
  \caption{Top panel: Density profiles of the models with CSM constructed in a
  way similar to \citet{moriya:17,moriya:18}, where it represents accelerating wind. 
  For comparison,
  black dashed line shows the model with low density wind only. 
  Bottom panel: Slopes of the corresponding CSM profiles 
  from the top panel as a function of radial
  coordinate.} 
  \label{fig:density_acc}
\end{figure}

In the top panel of Figure~\ref{fig:density_acc}, we show the density profiles of the models
considered in this section. As was mentioned in Section~\ref{models}, the high 
density CSM in these models is truncated at the radius $10^{15}\,{\rm cm}$. Above this
radius, we attach a low density RSG wind
with the mass loss rate $\dot{M}=2.0\times10^{-6}\,M_{\odot}\,{\rm yr}^{-1}$.
For comparison, the black dashed line in the top panel of Figure~\ref{fig:density_acc} shows the
model with low density wind only. When constructing the models shown in 
Figure~\ref{fig:density_acc}, we imposed the condition of constant total CSM mass,
$M_{\rm CSM}=0.5\,M_{\odot}$, as in the previous section. 
Our motivation for it is based on the roughly equal amount of mass
needed to reproduce the early light curve of another SN, 2013fs, 
by the models including steady
($0.47\,M_{\odot}$, \citealt{morozova:17}) and accelerating 
($0.5\,M_{\odot}$, \citealt{moriya:17}) high density winds.
The bottom panel of Figure~\ref{fig:density_acc} shows the slope $n$ in the
CSM of the corresponding models from top panel, computed as
$n=-(d\rho/dr)r/\rho$, as a function of $r$. In this plot, one can see that in this
kind of CSM the density declines with radius rather steeply, 
with the values of $n$ comparable
to the ones from our Section~\ref{results_CSM} and larger in some regions
\footnote{However, the $\beta=5$ and $n=5$ models are not to be confused.}.

\begin{figure}
  \centering
  \includegraphics[width=0.475\textwidth]{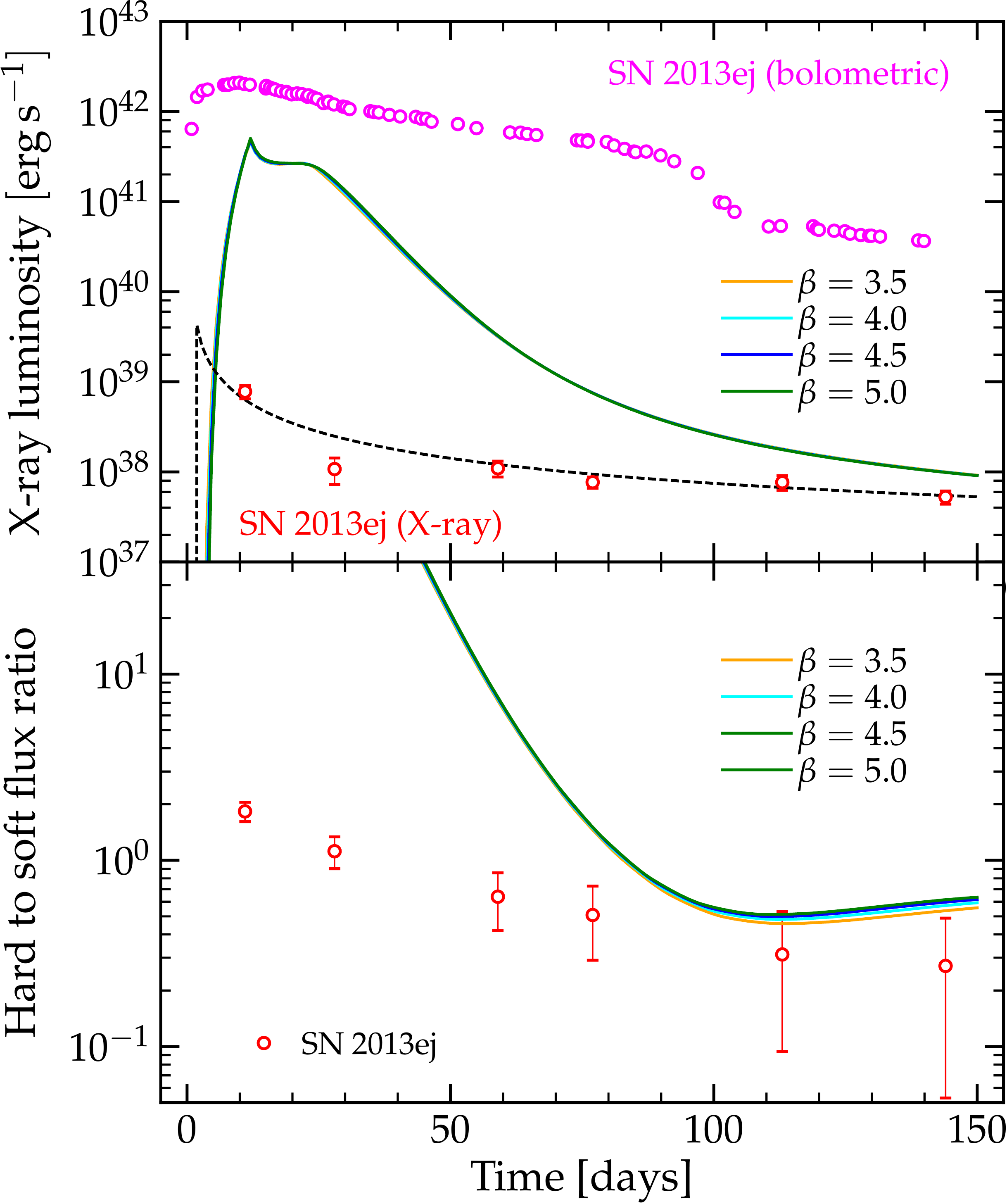}
  \caption{Top panel: X-ray luminosity obtained from the models shown in Figure~\ref{fig:density_acc}.
  The curves corresponding to the different values of $\beta$ are nearly identical.
  Black dashed line shows the X-ray signal from the model with low density wind only
  (mass loss rate $\dot{M}=2.0\times10^{-6}\,M_{\odot}\,{\rm yr}^{-1}$).
  For comparison, red markers show the luminosity of SN 2013ej in X-rays, and magenta
  markers show the quasi-bolometric luminosity of this SN derived from the optical observations 
  \citep{valenti:16}. Bottom panel: The ratio between hard ($2.0-8.0\,{\rm keV}$)
  and soft ($0.5-2.0\,{\rm keV}$) components of the obtained X-ray signal, compared
  to the values observed in SN 2013ej.} 
  \label{fig:lum_acc}
\end{figure}

In the top panel of Figure~\ref{fig:lum_acc}, we show the X-ray signal obtained from these models
using numerical approach outlined in Sections~\ref{athena} and~\ref{xspec},
and compare it to the observed X-ray signal from SN 2013ej. 
The final energy of the models is $0.68\,{\rm B}$ in all cases.
We transform the
flux into luminosity (multiplying by $4\pi$ times distance to 
SN 2013ej squared) and plot it together with the X-ray and quasi-bolometric luminosities
of SN 2013ej \citep[for the quasi-bolometric luminosity, see][]{valenti:16}. 
This plot shows that the X-ray flux we obtain from the models with
accelerating wind is few orders of magnitude larger than the observed X-ray flux
in the first $\sim50$ days, and its luminosity is almost comparable to the
quasi-bolometric luminosity of SN 2013ej. The light curves from the models 
with different values of $\beta$ are very
close to each other and nearly indistinguishable in the plot. For reference, the black
dashed curve shows the X-ray signal from the model with low density wind
of the mass loss rate $\dot{M}=2.0\times10^{-6}\,M_{\odot}\,{\rm yr}^{-1}$ only
(the same as blue curve in the top panel of Figure~\ref{fig:windden}). The
bottom panel of Figure~\ref{fig:lum_acc} shows the ratio between hard ($2.0-8.0\,{\rm keV}$) 
and soft ($0.5-2.0\,{\rm keV}$) components of the obtained X-ray signal, compared 
to the values observed in SN 2013ej.

\begin{figure}
  \centering
  \includegraphics[width=0.475\textwidth]{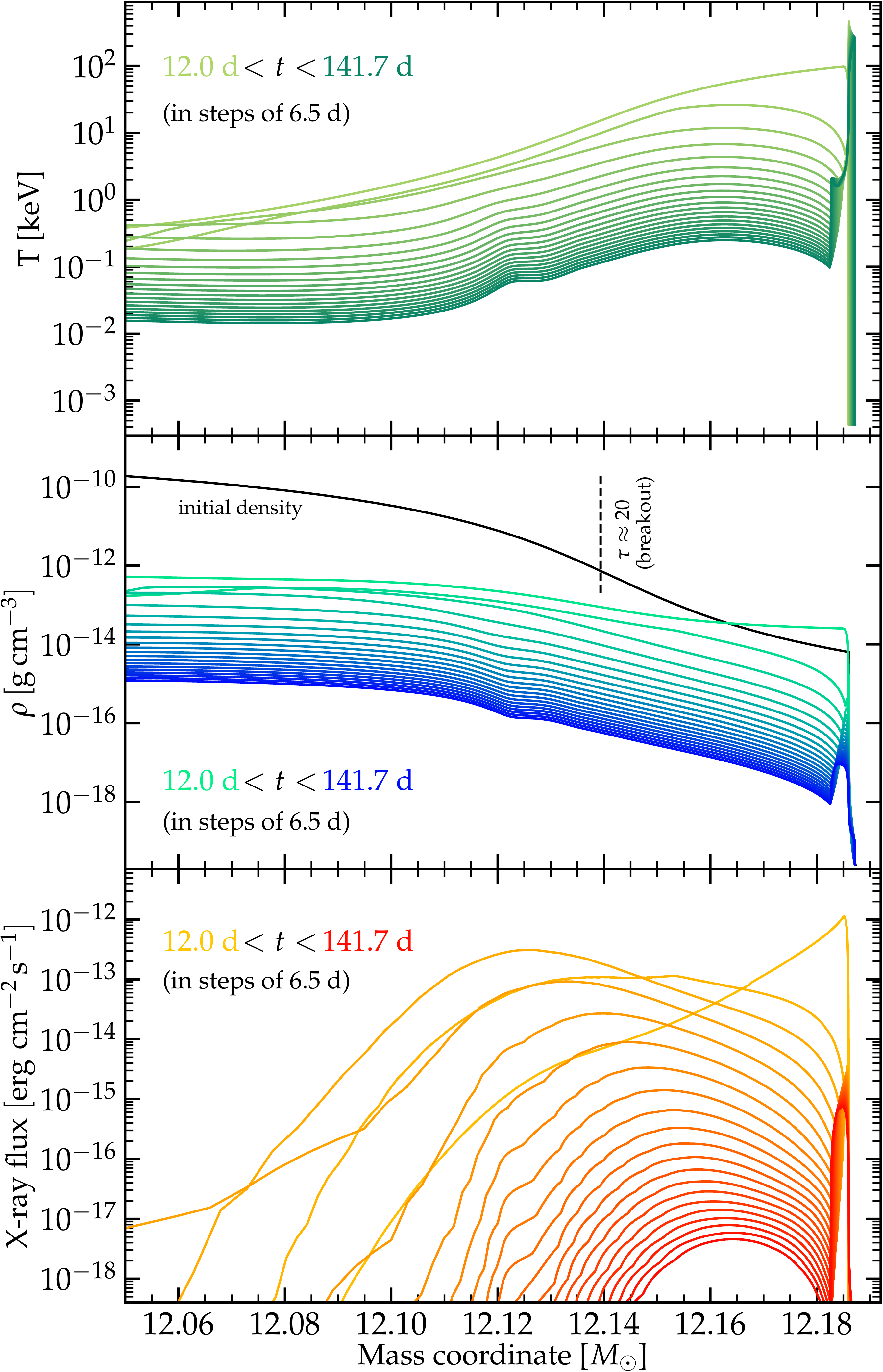}
  \caption{Evolution of the temperature profile (top panel), the density 
  profile (middle panel), and the X-ray emission from each grid
  cell (bottom panel) of the forward and reverse shock waves in
  $\beta=5$ model from Figure~\ref{fig:density_acc}.} 
  \label{fig:hydro_acc}
\end{figure}

In order to identify the source of excessive X-ray emission seen
in Figure~\ref{fig:lum_acc}, we plot the hydrodynamical evolution of the
$\beta=5$ model from Figure~\ref{fig:density_acc} in Figure~\ref{fig:hydro_acc}.
The top panel of Figure~\ref{fig:hydro_acc} shows the temperature evolution
in the outer $\sim$$0.15\,M_{\odot}$ of the CSM between days $12$ and $142$ 
(by the day $12$, the shock breakout in this model has already happened).
This plot demonstrates that the outer layer of this CSM gets heated by the
original post-explosion
shock wave to very high temperatures, of the order of a keV and larger.
Subsequently, after the shock wave 
has reached the boundary between the CSM and the low density wind, 
it gives rise to the usual forward/reverse shock wave pattern, similar to the
one from Figure~\ref{fig:hydro}, jammed along the
right hand side of Figure~\ref{fig:hydro_acc} (note different range
of the mass coordinate in these figures).

The middle panel of Figure~\ref{fig:hydro_acc} illustrates the evolution of
density in this model. For reference, we plot the initial density profile with
a solid black line. The velocity of matter ($v$) in the outer layer of this CSM is 
$\sim$$15,000\,{\rm km}\,{\rm s}^{-1}$ in our simulations, therefore we estimate
that the shock breakout in this model should happen at the optical depth 
$\tau\approx c/v \approx 20$, where $c$ is the speed of light. We calculate
the optical depth in the CSM by finding the integral 
$\tau=\int_r^{R_{\rm CSM}}\kappa\rho dr$, where the opacity $\kappa$ is equal to
$0.34\,{\rm cm}^2\,{\rm g}^{-1}$. In the middle panel of Figure~\ref{fig:hydro_acc}, 
we mark the location of shock breakout by a thin dashed line. To the right of this
line, we expect that the temperature of matter will be affected by the radiative
cooling during the shock breakout. However, to the left of this line, the radiation
will be trapped before the breakout, and the matter will be able to reach the
temperatures shown in the top panel of Figure~\ref{fig:hydro_acc}.

The bottom panel of Figure~\ref{fig:hydro_acc} shows total X-ray flux
from each numerical grid cell of the $\beta=5$ model. 
It demonstrates that in the first $\sim$$40-50$ 
days of the evolution the X-ray emission from this model is dominated by
the outer $0.1\,M_{\odot}$ of the CSM, due to its high temperature
and relatively high density. Note that in this plot we show the X-ray flux already after 
accounting for the absorption by all material outer to a given cell, and only in the
energy range $0.5-8.0\,{\rm keV}$. This explain very large X-ray emission
from this model seen in Figure~\ref{fig:lum_acc}.
After day $\sim$$50$, the reverse
shock wave reflected at the boundary between the CSM and the low
density wind becomes the dominant source of the X-ray signal. 
By the end of the simulation, at day $\sim$$150$,
the X-ray flux from this model almost entirely comes from the outer
shocked ejecta, as in the cases described in Sections~\ref{results_low} 
and~\ref{results_CSM}.

It is important to note that the X-ray emission seen in the bottom panel
of Figure~\ref{fig:hydro_acc} comes from the outermost region of the CSM, 
which quickly becomes optically thin (within the first few tens of days after
the breakout). 
Therefore, it cannot be captured by the codes (both gray and 
multigroup), which track the diffusion of radiation from the inner ejecta regions
either up to the photosphere, or to the point where the radiation decouples from the 
matter. In order to properly
describe this X-ray emission, one needs to look at the temperature
across the entire model, including the optically thin regions.

On the other hand, our numerical simulations use a range of assumptions,
which likely lead to the overestimate of the temperature in 
CSM. Perhaps, the most important of those is neglecting the radiative
cooling. In order to check how the
radiative cooling by X-rays may affect our results, we performed test simulations
similar to the one in Section~\ref{sec_cooling}, employing the cooling 
function $\Lambda$ from Equation~(\ref{cooling}). This formula estimates the total 
loss of energy due to free-free (dominates at the temperatures 
$T\gtrsim 2.6\times 10^7$) and bound-free
(dominates at the temperatures $T\lesssim 2.6\times 10^7$) X-ray emission 
\citep{chevalier:16}. In Section~\ref{sec_cooling}, we 
subtracted the entire amount of this energy from the shocked ejecta, 
without taking into account that part of it may be reabsorbed by the matter.
Nevertheless, the resulting X-ray light curve was only slightly affected by the cooling
(see Figure~\ref{fig:cool}).
However, the hydrogen column of the CSM in this section is much larger than
the hydrogen column of the shocked ejecta in Section~\ref{sec_cooling}, and part of the 
emitted energy will be inevitably reabsorbed back. 

To roughly account for the reabsorption, we multiply $\Lambda$ 
by the factor $\exp(-H/H_0)$, where $H$ is the hydrogen column at a given
point (see Section~\ref{xspec} for the definition) and $H_0$ is its characteristic 
value. Specifically, we apply the cooling function at each time step by adding the
term $-\exp(-H/H_0)\Lambda n_e n_i dt$ to the energy density in all grid cells,
where the temperature is higher than $5\times10^4\,{K}$. 
We perform the simulations with two values of $H_0$, $10^{22}$ and $10^{24}\,{\rm cm}^{-1}$,
and record the amount of energy extracted from the model in each case. 
After that, we analyse these simulations with \texttt{XSPEC} and compare the
obtained X-ray light curve to the extracted energy. This approach is not self-consistent, but
it helps us understand whether we subtract the right amount of energy from
the model or not.

\begin{figure}
  \centering
  \includegraphics[width=0.475\textwidth]{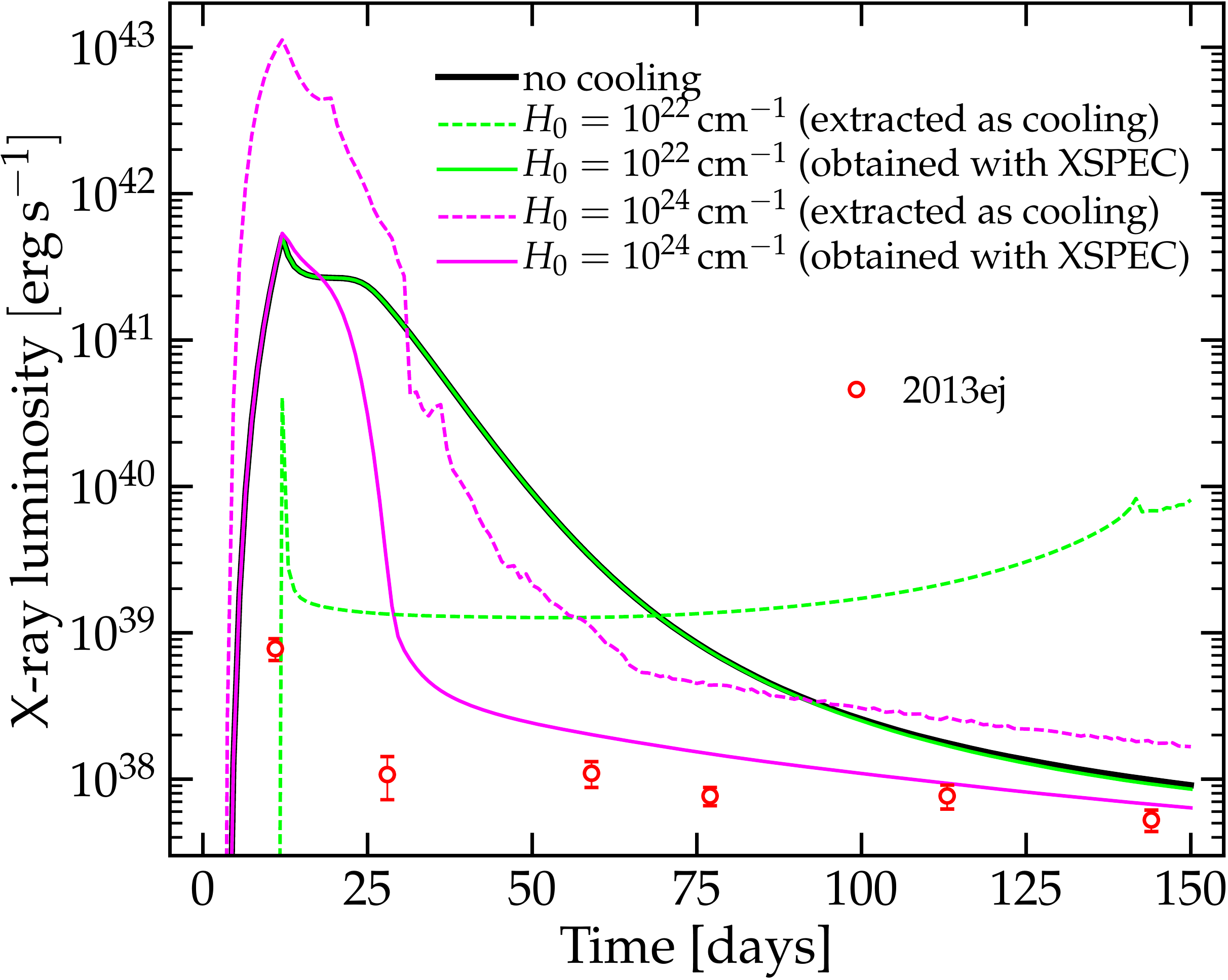}
  \caption{X-ray light curves obtained from the $\beta=5$ model with radiative cooling
  (green and magenta solid lines), compared to the total amount of energy
  extracted from the corresponding hydro simulations (green and magenta dashed lines).
  The black line shows the X-ray light curve from the model without cooling.
  The plot shows that in the case of $H_0=10^{22}\,{\rm cm}^{-1}$ we
  do not subtract enough energy from the model, while in the case 
  of $H_0=10^{24}\,{\rm cm}^{-1}$ we subtract more energy than necessary. We
  conclude that the correct X-ray light curve from this model likely lies between the
  solid green and solid magenta curves.} 
  \label{fig:columns}
\end{figure}

Figure~\ref{fig:columns} summarizes the results of this test for the $\beta=5$ 
model from Figure~\ref{fig:density_acc}. The black solid line shows the light curve obtained
without radiative cooling. Comparing solid and dashed green lines for the model
with $H_0=10^{22}\,{\rm cm}^{-1}$, we can see that our cooling method does not
remove a sufficient amount of energy in this simulation during the first $\sim$$70$ days, 
therefore, the X-ray light curve stays the same as in the case without cooling.
The fact that after this time the dashed green line is much higher than the
solid green line does not contradict to this, because the X-ray emission in that period is 
already dominated by the reverse shock wave, which is weakly affected by the cooling
(see Section~\ref{sec_cooling}), and the energy given by green dashed line come from 
the deeper regions, which do not contribute much to the signal anymore. 
Instead, comparing solid and dashed magenta lines for the simulation
with $H_0=10^{24}\,{\rm cm}^{-1}$, we can see that in this case we remove more energy from
the model than necessary, and the X-ray light curve does become weaker.
Nevertheless, it stays excessively large in the first $\sim$$30$ days. This suggests
that the correct X-ray signal from this model may lie between the solid magenta and
the solid green curves. However, this needs to be shown in more accurate numerical
models of the X-ray emission.

\begin{figure}
  \centering
  \includegraphics[width=0.475\textwidth]{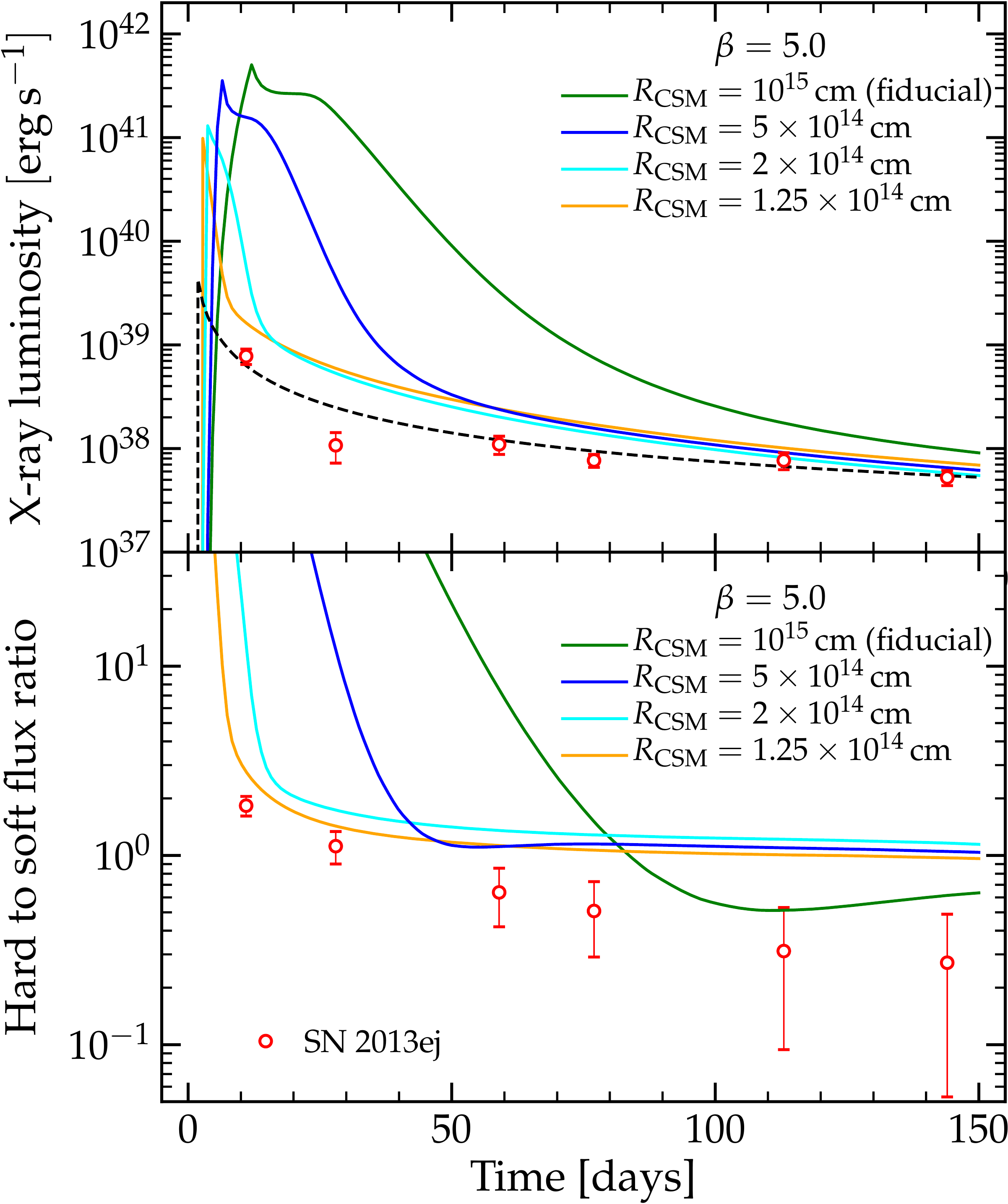}
  \caption{Top panel:  X-ray luminosity obtained from the $\beta=5$
  model truncated at the CSM radii of $R_{\rm CSM}=5\times 10^{14}\,{\rm cm}$,
  $2\times 10^{14}\,{\rm cm}$, and $1.25\times 10^{14}\,{\rm cm}$. 
  Black dashed line shows the X-ray signal from the model with low density wind only
  (mass loss rate $\dot{M}=2.0\times10^{-6}\,M_{\odot}\,{\rm yr}^{-1}$).
  For comparison, red markers show the observed X-ray luminosity of SN 2013ej.
  Bottom panel: The ratio between hard ($2.0-8.0\,{\rm keV}$)
  and soft ($0.5-2.0\,{\rm keV}$) components of the X-ray signal from the
  $\beta=5$ model with reduced CSM radii.} 
  \label{fig:truncated}
\end{figure}

Finally, in Figure~\ref{fig:truncated}, we check the dependence of our results on 
the size of the dense CSM with varying slope. For this test, we focus on the $\beta=5$
model from Figure~\ref{fig:density_acc} and truncate it at the radii 
$R_{\rm CSM}=5\times 10^{14}\,{\rm cm}$,
$2\times 10^{14}\,{\rm cm}$, and $1.25\times 10^{14}\,{\rm cm}$, where the latter
value is equal to the size of the best fitting constant slope CSM model from 
Section~\ref{results_CSM} ($1800\,R_{\odot}$). 
Above the truncated dense CSM, we attach the  
$\dot{M}=2.0\times10^{-6}\,M_{\odot}\,{\rm yr}^{-1}$ low density wind as we did before.
The top panel of Figure~\ref{fig:truncated} demonstrates that the excessive X-ray signal 
from the $\beta=5$ model is very sensitive to the size of the dense CSM, and it
nearly vanishes for the smaller valuer of $R_{\rm CSM}$.
In \citet{moriya:17,moriya:18}, the value of $R_{\rm CSM}=10^{15}\,{\rm cm}$ was
chosen somewhat arbitrarily, but the top panel of Figure~\ref{fig:truncated} suggests that the X-ray
signal from SNe~II, or the absence of such signal, may put a constraint on the size of the dense
CSM and, therefore, on the duration of its formation.


However, using the smaller CSM radii of the $\beta=5$ model,
we nevertheless cannot achieve a good agreement with the observed hardness of the
X-ray signal from SN 2013ej. The bottom panel of Figure~\ref{fig:truncated} 
shows the ratio between the hard and soft components of the X-ray flux, from which we
can see that the models overestimate the observed temperatures.
This can be naturally explained by the fact that trimming the CSM in the $\beta=5$ model
we make it more similar to the models from Section~\ref{results_CSM}. 
In those models, the X-ray signal,
instead of being dominated by the outermost overheated region of the CSM, mainly
comes from the reverse shock wave formed at the boundary between the dense CSM
and the low density wind. The results of Section~\ref{results_CSM} 
(see Figure~\ref{fig:windacc}) suggest that the hardness of
this X-ray signal depends on the velocity of the ejecta, which, in turn, is determined by the CSM
slope. And indeed, because the slope of the $\beta=5$ model reaches the values of $\sim$7 and
higher (Figure~\ref{fig:density_acc}), the hardness of its X-ray emission seen
in the bottom panel of Figure~\ref{fig:truncated} is remarkably similar to the 
hardness seen in Figure~\ref{fig:windacc}
for the models with $n=7$ and $n=8$. This once again emphasizes the importance of
the CSM slope in reproducing the right temperature of the observed X-ray signal. 
In addition, we obtained the optical light curve from the $\beta=5$ model trimmed
to the radius $R_{\rm CSM}=1.25\times 10^{14}\,{\rm cm}$ using \texttt{SNEC}, 
and found that it is almost identical to the ligth curve
from the $n=8$ model of Section~\ref{results_CSM}.

Due to the assumptions we use in our analysis, the results we obtain in this section by
no means can rule out the models of accelerating wind, or question their validity.
More detailed simulations, taking into account radiative cooling, Comptonization
of the optical photons, and incomplete equipartition between the ions and electrons, are
needed in order to have a more clear picture of the X-ray emission from 
these models. Nevertheless, this simplistic approach shows that the early 
SN~II observations in X-ray can play an important role in constraining the 
properties of this kind of CSM. Even is the excessive heating we see in these
models is entirely eliminated by the radiative cooling, the question of whether
the outer layers of this CSM are accelerated to the right velocities in order to
ensure the right temperature in the reverse shock wave 
(and, consequently, correct X-ray spectrum) still stays
open. After all, if the models including dense CSM with varying slope
(aka accelerating wind) indeed represent the right way to describe the 
pre-explosion configuration of SN~II progenitors, more accurate numerical
simulations of the X-ray signal from these models will confirm it
by reproducing the early X-ray data observed in SNe~II.


\section{Conclusions}
\label{conclusions}

It goes without saying that a good SN model should reproduce observations
across the entire electromagnetic spectrum, and for the core-collapse SNe~II also in 
neutrinos \citep{murase:17,seadrow:18} and gravitational waves 
\citep{morozova:18}. This is a tremendously difficult problem, and the
presence of a dense CSM of unknown nature makes it even more complicated.
In this study, we attempted to find a model that could fit both optical photometry
of SN 2013ej until the end of the plateau phase and its
X-ray signal in the first $\sim$$150$ days. In order to accomplish this, we
had to use a set of simplifying assumptions. Here,
we summarize the conclusions of this work, which we think are robust.

\begin{itemize}
\item Inclusion of a dense CSM in the pre-explosion model changes the location
of the forward/reverse shock waves responsible 
for most of the X-ray signal in SNe~II. Instead of
forming at the interface between the RSG surface and the low density wind, they form
at the interface between the dense CSM and the low density wind. In this
scenario, the reverse shock wave probes the outer edge of the dense CSM, while
the forward shock wave propagates into the regular low density 
RSG wind. This may explain the fact that radio observations of SNe~II favor
low mass loss rates from their progenitors, similar to the regular mass loss rates of
RSGs \citep{chevalier:06}.

\item The density slope in the CSM sets up the conditions for the acceleration of the
original post-explosion shock wave. The steepness of this slope defines the final
velocity of the outermost CSM layers, which, in turn, determines the temperature
in the reverse shock wave and the observed X-ray spectrum. Steeper CSM profiles
result in a harder X-ray emission. Therefore, while modeling SN optical photometry helps
constraining the total CSM mass, modeling their X-ray signal may help constraining
the CSM slope.

\item We find that $\rho\propto r^{-2}$ (dense steady wind) CSM profile
cannot accelerate the original shock wave enough to explain the ratio between
the hard and soft X-ray components observed in SN 2013ej. More accurate treatment
of radiative cooling would likely make the agreement worse. Instead, we show that a 
steeper, $\rho\propto r^{-5}$ CSM with the same total mass and radial extent 
does better job in
reproducing optical and X-ray emission from SN 2013ej (see Figure~\ref{fig:fit} for the 
full set of parameters in this model).

\item On the other hand, we find that steeper and more extended CSM profiles, similar
to the $\beta$-law models of accelerating wind, may create conditions that lead to the excessive
X-ray emission, which is not observed in SNe~II. However, this conclusion is sensitive
to the assumptions that we use and must be carefully checked by more detailed simulations.
Nevertheless, we argue that reproducing X-ray signal from young SNe~II is an important
test that any viable CSM model should pass.
\end{itemize}

All simulations performed here assume spherical symmetry, but 
this assumption does not necessarily hold true in real SNe. In fact, the unusually strong
polarization seen in the spectra of SN 2013ej \citep{leonard:13,kumar:16} 
likely indicates a significant asymmetry
in the ejecta. Due to their complexity and computational cost, multi-dimensional numerical 
simulations of aspherical CSM interactions are still not very common.
Motivated by nebular observations of SN 1987A, \citet{blondin:96} performed two- and 
three-dimensional simulations of a shock wave entering an axisymmetric CSM with the
polar angle dependent density profile having minimum at the pole and 
maximum at the equator. They found that for a highly aspherical CSM
the shock wave forms a protrusion along the
axis up to $\sim$4 times larger than the main radius of the shock. Another approach was taken in the
works of \citet{couch:09,suzuki:16,afsariardchi:18}, 
where the authors considered a spherically symmetric density profile, 
but employed an aspherical explosion
mechanism to launch the shock wave. Under certain conditions, such setup may lead to the 
formation of an `oblique' shock wave propagating along the stellar surface 
in the direction of the polar angle towards the equator. 
\citet{jun:96} studied the enhancement of the Rayleigh-Taylor 
instability in a shock wave interacting with a
clumpy CSM and suggested that the presence of clumps can lead to a 
stronger X-ray and radio signal.

In addition to possible asphericity of the 
density profile or the explosion mechanism, 
a spherically symmetric shock wave accelerating in a steep density 
profile may become unstable with respect to small linear perturbations \citep{goodman:90,chevalier:90}. 
For example, \citet{sari:00} found that in the case of an adiabatic equation of state 
with $\gamma=5/3$ the minimum density slope at which the 
accelerating shock wave becomes unstable
is $n_c\approx 7.7$. This fact could be especially relevant for the CSM models 
from Section~\ref{results_Mor} with 
small values of $\beta$, where the slopes may reach values much higher than $n_c$
(see Figure~\ref{fig:density_acc}). Without detailed multi-dimensional numerical simulations it is very
hard to say how the effects of asphericity, clumpiness, and shock wave instability will influence the
results of our study. In the future, we hope to be able to perform such simulations and
address this question.

This study focused on a single well-observed object, but analysis of a larger sample
of X-ray bright SNe~II would give us much better idea on the variety of CSM around
RSGs and could potentially help clarifying its origin. In this view, we encourage 
observational programs of young SNe~II using current and future X-ray facilities 
\citep{chakraborti:12,chakraborti:13,chakraborti:16,ross:17}
and anticipate exciting results that will be obtained with those data.

\acknowledgments
We acknowledge helpful discussions with Adam Burrows, David Radice, Tony Piro
and Stefano Valenti. J.S. acknowledges support from NSF grant AST-1715277.
We thank our anonymous referee for his/her detailed and useful comments.

\bibliographystyle{apj}

\end{document}